**Onset of a conceptual outline map to get a hold on the jungle of cluster analysis**

Iven Van Mechelen
University of Leuven

Christian Hennig
University of Bologna

Henk A. L. Kiers
University of Groningen

**Abstract**
The domain of cluster analysis is a meeting point for a very rich multidisciplinary encounter, with cluster-analytic methods being studied and developed in discrete mathematics, numerical analysis, statistics, data analysis, data science, and computer science (including machine learning, data mining, and knowledge discovery), to name but a few. The other side of the coin, however, is that the domain suffers from a major accessibility problem as well as from the fact that it is rife with division across many pretty isolated islands. As a way out, the present paper offers a thorough and in-depth review of the clustering domain as a whole under the form of an outline map based on an overarching conceptual framework and a common language. With this framework we wish to contribute to structuring the clustering domain, to characterizing methods that have often been developed and studied in quite different contexts, to identifying links between methods, and to introducing a frame of reference for optimally setting up cluster analyses in data-analytic practice.

## 1. INTRODUCTION

By way of a working definition, cluster analysis can be defined as the collection of all possible methods to infer a priori unknown groups from data. This collection is immensely large, as illustrated by the following enumeration of names of a small subset from this collection (in alphabetical order, with for each method a single reference): additive clustering (Shepard & Arabie, 1979), alternative clustering analysis (Bailey, 2014), blockmodeling (Arabie et al., 1978), co-clustering (Govaert & Nadif, 2013), density-based clustering (Kriegel & Pfeifle, 2005), ensemble clustering (Singh et al., 2010), fuzzy clustering (Bezdek et al., 2005), grid-based clustering (Cheng et al., 2014), hierarchical clustering (Johnson, 1967), kernel-based clustering (Xu & Wunsch, 2009), *k*-means (Steinley, 2006), latent class analysis (Vermunt & Magidson, 2004), mixture modeling (McLachlan & Basford, 1988), model-based clustering (McNicholas, 2016), multipartitioning (Vichi, 2016), network clustering (Parthasarathy & Faisal, 2014), neural network-based clustering (Xu & Wunsch, 2009), partitioning around medoids (Kaufman & Rousseeuw, 1990), pyramidal clustering (Diday, 1986), rough set clustering (Düntsch & Gediga, 2016), semi-supervised clustering (Agovic & Banerjee, 2014), spectral clustering (Liu & Han, 2014), stream clustering (Aggarwal, 2014a), symbolic data clustering (Brito, 2016), time-series clustering (Caiado et al., 2016), uncertain data clustering (Aggarwal, 2014b). Moreover, many of these names refer to a family of clustering methods rather than a single such method.

The domain of cluster analysis shows a mixed picture. On the positive side, firstly, the domain is a meeting point for a very rich multidisciplinary encounter: Cluster analysis and cluster-analytic methods are being studied and developed in discrete mathematics (including graph theory, and (fuzzy) set theory), numerical analysis (including numerical and combinatorial optimization), statistics (including statistical modeling and computational statistics), data analysis and data science, and computer science (including machine learning, data mining, knowledge discovery, artificial intelligence, and pattern recognition), to name but a few.

Secondly, interest in cluster analysis is unremitting. This can be evidenced by a search in Web of Science (Core collection, July 2023), inspired by a somewhat similar search by Murtagh and Kurtz (2016) in Google Scholar: 28,243 papers appeared with “cluster analysis” in “All fields” and with publication date between January 1, 2018 and December 31, 2022, on average over 5,600 publications per year.

Thirdly, clustering goes with applications in very many areas of science, including (but certainly not limited to) astronomy, business and marketing, climatology, social and communication sciences, geochemistry, image processing, linguistics, medicine, plant and animal ecology, psychology, social sciences, and systems biology.

But the domain is also cursed with problems. In particular, the available clustering methods, models, and data-analytic techniques form an inconvenient and intricate jungle. Reasons for this include that they stem from many different areas, with different languages, concepts, priorities and concerns. Moreover, there are only a few bridges between these areas, with work in one area typically being hardly cited in the others. This hampers the identification of relations between methods developed in different fields, and goes with the fact that methods developed in certain areas are often “reinvented” in other ones (leading to different names for exactly or essentially the same method). What is perhaps most problematic is that there is no overarching conceptual framework to structure the clustering domain as a whole, to characterize the different families of methods that are part of it, and to deeply understand their interrelations. The above constitutes a major problem for methodological researchers of clustering methods, for students who want to familiarize themselves with the domain and may easily get lost in it, and for applied researchers on the lookout for suitable clustering methods for substantive problems.

As a way out, the present paper offers a thorough and in-depth conceptual review of the clustering domain. This review takes the form of an outline map based on an overarching conceptual framework. To be sure, many reviews of the clustering domain are already available (e.g., Aggarwal & Reddy, 2014; Everitt et al., 2011; Ezugwu et al., 2021; Gan et al., & Wu, 2020; Gao et al., 2023; Hennig et al., 2016; Jaeger & Banks, 2023; Jain, 2010; Rokach, 2010; Van de Velden et al., & Markos, 2019; Xu & Tian, 2015; Zhang et al., 2023). Most of these, however, have been organized in terms of surface characteristics and fragmentary subdivisions or have been based on a single perspective only, and involve a necessarily subjective selection of a subset of clustering methods to keep such reviews feasible. Our approach will be fundamentally different, in that our conceptual review will be rooted in an in-depth formal analysis of the clustering work flow. This will provide a frame of reference with well-wrought subdivisions guided by first principles and from multiple perspectives. Ultimately, it will address fundamental questions that contribute to structuring the clustering domain and to making it more accessible, to characterizing clustering methods that have often been developed and studied in quite different contexts in terms of their essential core properties, to identifying key links between them along with their most fundamental common and distinctive features, and to supporting users in optimally setting up and evaluating cluster analyses in data-analytic practice. To be sure, given the vast proportions of the challenge we are addressing, our attempt can only yield an *onset* of the frame of thinking we ultimately aim at. This implies that we will primarily provide the reader with a number of handles, without elaboration of all possible details.

A key insight at the basis of our formal analysis is the acknowledgement that there is no unique "best" clustering of a data set: The same data allow for various clusterings that can be justified in various ways. Clustering cannot be fully automated: User input and decisions are required to define what kind of clustering is required in any given situation. There is neither an agreed upon definition of what is a cluster nor a general formal definition of the clustering problem. This does not mean that clustering problems cannot be defined; rather, any definition has to be context-dependent and case-specific (see also Anderberg, 1973; Hand, 1994; Hartigan, 1985; Hennig, 2015; von Luxburg et al., 2012), and our outline map is meant to help with defining it sensibly in a given situation.

As target audience for the present paper we hope to reach, first of all, formal, methodological, and theoretical researchers from the many disciplines in which cluster analysis and cluster-analytic methods are being studied and developed (including both established scholars and students in these disciplines). Secondly, we also hope that applied researchers might benefit from reading this paper.

In the remainder of this paper we will first present an overall scheme of the data-analytic process as backbone of the outline map. Subsequently, we will enter into an in-depth discussion of the six constituents of the backbone and the most important taxonomic distinctions within each of them. Next we will give a brief, proof-of-concept preview of how the outline map could be used in cluster-analytic practice.

This paper further comes with two major sources of reference materials under the form of annexes. The first of these is a glossary (Appendix A). Its primary role will be to act as an explanatory dictionary with sharp definitions of terms. This is intended to make the paper self-contained and may act as the basis for a common language to support communication across boundaries of disciplines and to avoid miscommunication. The second of these is a checklist with six parts. Those correspond to the six constituents of the backbone of our to be proposed outline map (Appendix C). The checklist may be a most useful tool in guiding or evaluating a cluster-analytic process in data-analytic practice. It also may contribute to an in-depth characterization and understanding of (both existing and novel) clustering methods and their interrelations.

## 2. OVERALL SCHEME OF DATA ANALYSIS AS BACKBONE OF OUTLINE MAP

In Figure 1 we present an overall scheme that may act as the backbone of the outline map to be introduced. This scheme is based on a global account of a data-analytic process. As such, it links up with repeated calls to invest in “developing a more robust formal understanding of [this] process” (Peng & Parker, 2022, p. 18). It bears some similarity to process accounts of machine learning, including in particular the “Cross Industry Standard Process for Data Mining” or CRISP-DM (see, e.g., Wirth & Hipp, 2000). (A supplementary figure in which all components of the scheme have been filled out with specific contents based on the example to be analyzed in Section 4.1 below, is presented in Appendix B.) Whereas the focus of the outline map is on cluster analysis, the scheme refers to the broader concept of data analysis. This is because several aspects of a clustering process may go beyond clustering in the narrow sense. As an example, user concerns may include concerns other than finding or constructing groups. As another example, the data analysis in a cluster-analytic process may go beyond analyses that yield clusterings only.

As appears from Figure 1, the outline map we will offer for getting a hold on the clustering domain will comprise a series of six interrelated angles. From each of these angles one may characterize and evaluate cluster-analytic methods and processes (in a somewhat similar spirit as in ontologies of data mining: see, e.g., Panov et al., 2008). For this purpose, we will advance within each of the angles a set of conceptual specifications and taxonomic distinctions. In general, the categories within the resulting taxonomies will not be mutually exclusive; furthermore, in some cases, they will be intended to be exhaustive. Importantly, angles and categories are meant to remind the user of the different aspects that are part of the clustering process and the different perspectives from which they can be considered.

The left box in Figure 1 contains, on the one hand, goals and, on the other hand, data as key inputs for the data-analytic process (D’Agostino McGowan et al., 2022; Hand, 1994). In our further elaboration we will especially focus on the steering aspects of these inputs. The right box represents the constituents of the data-analytic process in the broad sense, which we split into three phases (the boundaries between which are not very sharp: see below). The first of these (i.e., the input preprocessing) pertains to the preparation of the data, and the last (i.e., the output postprocessing) to operations on the data-analytic results for presenting and interpreting them; the phase in between the two will further be called the *core data analysis* (not to be confused with the same term used by Mirkin, 2019, with a different meaning). In our elaboration we will especially focus on the descriptive aspects of these three phases.

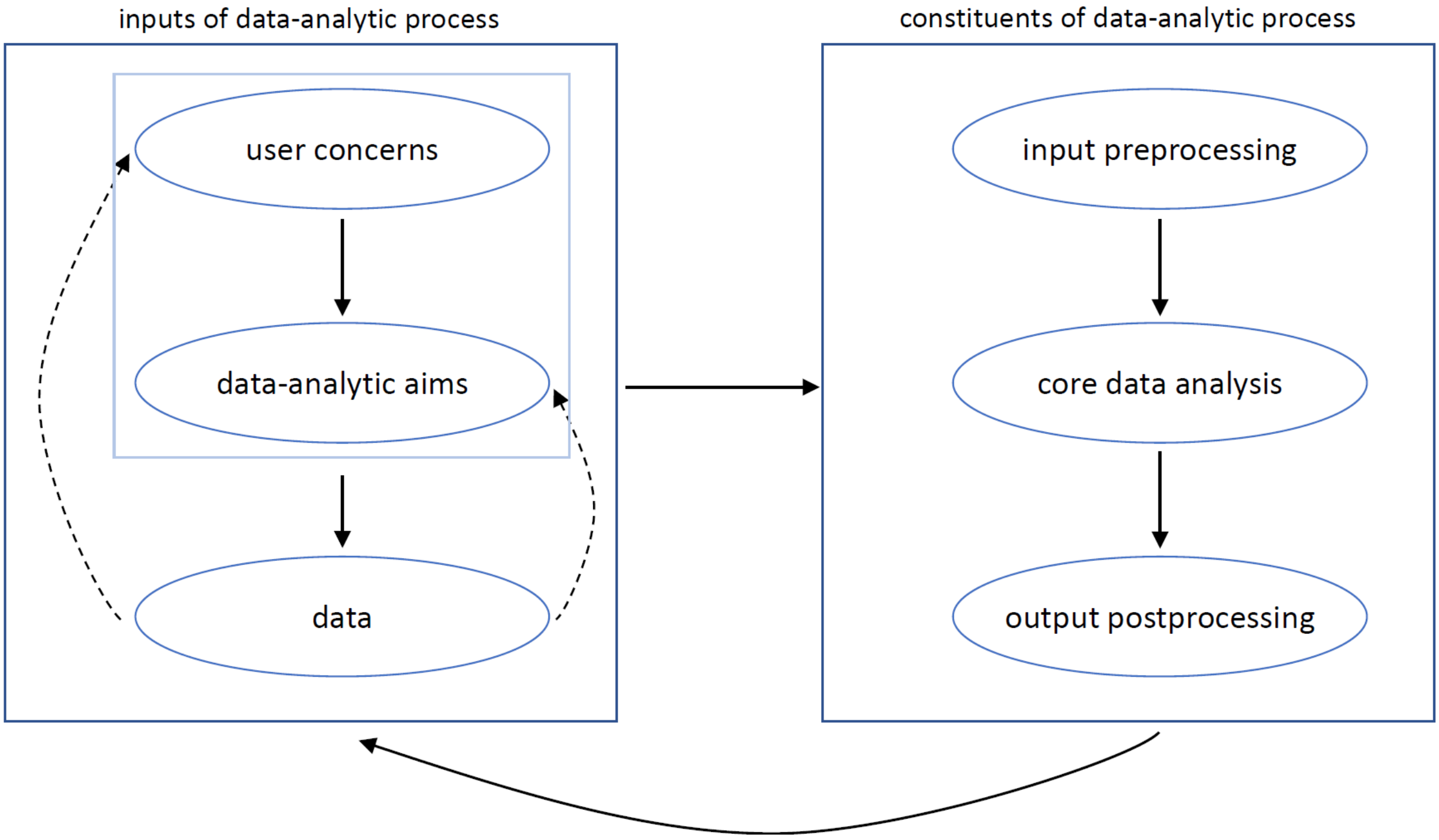

*Figure 1.* Overall scheme of data analysis as backbone of outline map

The whole of the arrows between the boxes represents the inherently dynamic and iterative nature of the data-analytic process, which often will proceed through multiple cycles. Indeed, one may start something, but subsequently other things may come along the road (e.g., alternative possibilities for analyzing the data that go with alternative concerns/aims). Incidentally, consultancy processes in data analysis and statistics often are similarly dynamic and iterative in nature.

Furthermore, the arrows between as well as within the boxes represent logical and/or temporal order, input-output type of relations, and relations of the type "has to be translated into". As an example, user concerns (including user needs and substantive domain or application-based research questions) contain critical information for determining what should be desirable output from the data-analytic process. Therefore, also the many choices that are to be made in that process need to be informed by the user concerns. (This links up with what is referred to in artificial intelligence as the alignment between the capabilities of a learning system and the human user's goals and preferences: see Vamplew et al., 2018). That is why they are given logical priority in the scheme and why they should, ideally speaking, also have temporal priority. User concerns further have to be translated into data-analytic aims. "Translation" is used here in more than a metaphorical sense, as user concerns and data-analytic aims are typically being formulated in different languages or types of speech (see also Hennig, 2010). As a second example, the whole of user concerns and data-analytic aims constitutes important input for choices to be made at the level of the data to be collected (in terms of data type, study design, etc.), in order to enable the researcher to meet the wishes in question.

We conclude this section with five important remarks. Firstly, in practice, not seldom situations show up that deviate from the temporal/logical order put forward in our scheme; we do not want to dismiss these. Examples include that initial user concerns or substantive domain questions may be (very) vague, or hard to elicit, or not answerable, and that data may contain information that had not been anticipated in advance and that may lead to a change of aims or analysis choices. All this is why the scheme contains dotted arrows denoting input from the data for sharpening user concerns and data-analytic aims. Similar (albeit less prominent) feedback loops would be possible at other places, too, with, more in general, iterative cycles also occurring within the left and right boxes of the scheme.

Secondly, the arrows in most cases do not signify one-to-one relations. As an example, the left box with the data-analytic input most often can be translated into multiple possible specifications of the data-analytic process. Similarly, in most cases there is not a one-to-one relation between user concerns and data-analytic aims, although, when facing a particular user concern, it is a legitimate question to look for a corresponding data-analytic aim and vice versa.

Thirdly, in practice, a number of aspects and taxonomic distinctions could be located within multiple possible ellipses of the scheme. As an example, one may think of the data-analytic aspect of variable selection which, within a clustering context, could be taken care of in the stage of input preprocessing (in so-called filter methods) as well as in the stage of the core data analysis (in so-called wrapper methods). When discussing each of the ellipses more in detail in the next sections of this paper, we will, for simplicity's sake, discuss such aspects within the context of a single ellipsis only. In a different but somewhat related vein, within many ellipses of our scheme, there may be elements corresponding to elements in another ellipsis. As an example, one could consider many particular data-analytic aims that correspond (provided some rewording) to particular descriptive characteristics of the core data analysis, and vice versa. In our more detailed discussion below of each of the six ellipses in Figure 1, we will, however, only list one of the corresponding elements and avoid doubling, for simplicity's sake.

Fourthly, whereas the iterative, multi-cycle nature of the data-analytic process as represented by the scheme of Figure 1 allows users to take advantage of unanticipated insights and to address follow-up questions, it may also imply a considerable inferential challenge. In particular, it can lead to biased conclusions and unwarranted overconfidence when data-analytic operations in some cycle are de facto chosen based on the outcome of a previous cycle, whereas their theoretical justification requires that this is not the case (see also Gelman & Loken, 2014).

Fifthly, a critical point of departure in the present paper is that all six angles or boxes as distinguished in Figure 1 are necessary to fully grasp a cluster-analytic process (in view of guiding or evaluating such a process in data-analytic practice). Furthermore, all final five angles are considered essential for a full characterization and a deep understanding of clustering methods and their interrelations (leaving aside the first angle which rather pertains to the substantive context in which the methods are applied). One might wonder whether all this is the case, indeed, and whether, for instance, focusing on the core data analysis only would not suffice for a proper account of methods of cluster analysis. There are several reasons why all angles are needed. To start with, the goal of an action or object –in our case: the data-analytic aim(s) behind a clustering method– is indispensable for understanding the function of that action or object. For any method it is further of key importance to know the type and nature of the input on which that method operates (or may operate), and any requisites it may imply on these; for a data-analytic method, the data constitute the most critical part of this input. Finally, preprocessing of a method's input and postprocessing of its output are inevitable as well: For some methods certain operations of pre- or postprocessing are mandatory (i.e., they are inextricably bound up with the methods in question), depending, among other things, on the nature and the type of the data at hand. Moreover, the same type of pre- or postprocessing operations may for certain methods require separate steps before or after the core data analysis, whereas for other methods they are actually part of it; this implies that the boundaries between pre- and postprocessing on the one hand and core data analysis on the other hand are semipermeable.

## 3. CONSTITUENTS OF BACKBONE

### 3.1 User concerns

We define the user as the person (or the group of persons, company, etc.) with a substantive domain interest at the basis of the data-analytic process. This may be a substantive or applied researcher, a customer of a data-analytic consultant, or a person employed in any field of practice. User concerns refer to what is important to the user, what the user aims at, what the user needs, wishes, or hopes to get ultimately from the data analysis. These may further be associated with substantive domain or application-based research questions.

Table 1 presents a number of taxonomic distinctions with regard to user concerns. We intend these distinctions to be as exhaustive as possible, in that they should allow for a characterization of all possible such concerns.

The table lists two major types of concerns: general and grouping-related concerns, with the latter, unlike the former, pertaining to learning or inducing unknown groups and their characterization from the data. Given the focus of the present paper on cluster analysis, one may wonder why we would not start with grouping-related concerns or even limit ourselves to such concerns. The reason for this is that, in practice, three different possible scenarios may occur: Firstly, the primary interest of the user may be in finding groups (leaving aside the possibility of additional, secondary interests of that same user). Secondly, the user may have a primary interest other than finding groups, but for some reason this may go with a secondary interest in finding groups. As an example, one may think of a user who, as primary interest, wants to predict for each member of some population of patients the survival time, but who, as a secondary concern, wants to take into account that as yet unknown subpopulations may exist that go with different prediction rules. Thirdly, finding groups may simply not figure among the original user concerns; the need to find groups might then nonetheless show up later in the data-analytic process. To illustrate, one may revert to the example of predicting survival time, but now without the initial secondary wish to take possible heterogeneity of the patient population into account, while nevertheless such heterogeneity might come to light later on in the data-analytic process. Conversely, a user may also originally believe that clusters are needed, whereas they are in fact not, with the consequence that the initial concern of finding groups may subsequently vanish during the data-analytic process.

Table 1
*Taxonomic distinctions with regard to user concerns*

| **General concerns** | **Grouping-related concerns** | |
|---|---|---|
| *Content* | *Nature* | *Relative importance* |
| explanation | finding groups | primary original concern |
| identification | constructing groups | secondary original concern |
| prediction | | not among original concerns |
| preparation of action | | |
| information compression | | |
| exploration | | |
| evaluation | | |

We now turn to a more detailed account of the user concerns listed in Table 1. Within the class of **general concerns**, a distinction can be drawn on the basis of the contents of the concerns:

- *Explanation*: This may pertain to uncovering the causes underlying the phenomenon of interest, in terms of proximal (i.e., immediate) precursors or of more distal factors at the basis of the phenomenon's genesis; relatedly, it may also pertain to uncovering the mechanisms and processes underlying the phenomenon in question. As an example, one may refer to questions on etiology or pathogenesis in medicine.
- *Identification*: pertains to the answer to questions of the type "what is this", that is to say, questions on the identity of the objects of interest. As examples, one may think of diagnostic questions in medicine and object recognition questions in image analysis.
- *Prediction*: pertains to any type of forecasting. To be sure, as a matter of conceptual precision, we limit the concept of prediction here as limited to direct prediction, while excluding indirect predictions over a chain of steps in the further future. The latter may be exemplified by a search for an etiological explanation for a disease, in view of devising in the future treatments of it that could then be predicted to be more successful; we typify this here in terms of the aim "explanation" rather than "prediction".
- *Preparation of action*: pertains to the fact that the user intends to do something (e.g., setting up a production facility, or launching a marketing campaign) and wants to (optimally) prepare for this.
- *Information compression*: one may wish to reduce an abundance of information, either for oneself or for communication to others. Such a compression concern involves issues of information loss and information preservation, including how much information can be derived from the compressed information (Needham, 1965), and including the question of whether the information that is preserved is good or sufficient for the aim for which it is to be used.
- *Exploration*: this can be defined as the concern to discover unexpected elements or patterns. Exploratory approaches are often put in contrast to confirmatory approaches or endeavors to test a priori specified expectations, hypotheses, or theories. However, as implied by the term "unexpected", exploration typically will not take place in a theoretical vacuum.
- *Evaluation*: the user may wish to evaluate various aspects. As an example, (s)he may wonder whether some intervention yielded a good effect. Any evaluation requires a specification of what is to be considered "good" versus "bad" (i.e., an evaluative standard). In general, the user or domain expert is uniquely qualified to provide such a standard.

Different types of general concerns can further be combined. As an example, one may wish to go for better founded predictions by simultaneously looking for suitable, associated causal explanations.

Within the class of **grouping-related concerns**, two taxonomic distinctions can be drawn. The first of these pertains to the difference between *finding* and *constructing* groups, and the second to the relative importance of grouping-related concerns compared to other data-analytic concerns.

Regarding the **first** taxonomic distinction, "*finding groups*" refers to the aim that clusters resulting from the analysis should correspond to a meaningful underlying distinction: Objects in a cluster should have something meaningful in common, beyond their observed data values, that distinguishes them from objects that do not belong to the same cluster. This idea is nicely expressed by Socrates/Plato as "dividing things again by classes, where the natural joints are, and not trying to break any part, after the manner of a bad carver" (Plato, approx. 370 BC). An example would be a cluster of pixels belonging to streets, say, in image segmentation, distinguished from other objects in the image. "*Constructing groups*" means that clustering is carried out because clusters are required for practical, pragmatic reasons (e.g., data compression or assignment of a number of action alternatives), whereas an interpretation in terms of an unobserved underlying truth is not relevant (see also Hand, 1998). –Incidentally, recent literature on diagnosing “clusterability” of data sets (e.g., Adolfsson et al., 2019) is concerned with “finding groups”, but has no implications on “constructing groups”.– One may note that the contrast “finding versus constructing” is not quite as clearcut as it might appear at first sight: On the one hand, even if finding an unobserved meaningful distinction is not a primary aim, in many situations it can be useful to have clusters in line with such meaningful distinctions. On the other hand, even if “finding” rather than “constructing” is declared as a major aim, underlying meaningful distinctions cannot be identified from data without further assumptions and specifications regarding how supposedly "true" underlying categories of interest relate to the observed data. Arguably, the data-analytic operations involved in computing a clustering in any case construct a classification on the data, regardless of whether the user is interested in "finding" or "constructing". There is no guarantee that any clustering computed from data corresponds to meaningful underlying categories. A user interested in finding meaningful structures will not only need a clustering, but also a convincing connection to relevant subject matter knowledge and in most cases also a data-based cluster validation (see further Section 3.6). Corresponding to the practical difficulty to ensure the meaningfulness of a cluster-analytic "finding", the philosophical theory of constructivism states that even what we think of as "meaningful categories" are constructed by the human processes of perception and cognition, whereas philosophical realists hold that such categories exist independently of the human observers. However, even a constructivist may accept "finding of meaningful structure" as a legitimate concern, as long as it is acknowledged that the assignment of underlying meaning to found clusters by the user is still a constructive act (see Bird and Tobin, 2023, for an in-depth account of this discussion regarding "natural kinds", and Watson, 2023, for connecting it to applied unsupervised learning from data).

Regarding the **second** taxonomic distinction within the class of grouping-related concerns, initial user concerns may include either general concerns or grouping-related concerns or both. As an example of the latter, one may recall the case of predicting patient survival time, combined with the wish to take a possible heterogeneity of the patient population into account (= combination of general concern of prediction with grouping-related concern). As a second example, one may think of evaluating the effect of an intervention to reduce customer churn, while assuming that there may be differential effectiveness for different types of customers (= combination of general concern of evaluation with grouping-related concern). The second taxonomic distinction can be drawn on the basis of the timing and the relative importance of the grouping-related concern(s): primary, secondary, or not among the original concerns. A grouping-related concern is called primary if it figures among the initial concerns, either as the only concern or in combination with (one or more) general concerns, but in the latter case with a degree of importance that is greater than or equal to the importance of the general concern(s). The categories “secondary” and “not among the original concerns” are defined accordingly.

Combinations of general and grouping-related concerns will often go beyond a mere co-occurrence of the two types of concerns. As an example, constructing groups may be used as a *means* to meet some general concern, such as information compression or data exploration.

Finally, we emphasize that the user concerns listed here regard the formulation and definition of the data-analytic problems of interest, but not their solution. We assume that users have a genuine and open-minded interest in what the data can contribute to address their problems rather than, for example, aiming at confirmation of their prior beliefs, or achieving statistical significance for increasing chances of publication. We are aware that in reality this assumption may not always be fulfilled, and that consequently there may be skepticism regarding too much influence of the users and their concerns on the analysis. We can only acknowledge the tension between the importance of user concerns for choosing and carrying out appropriate analyses (also emphasized by von Luxburg et al. 2012), and the danger of bias caused consciously or unconsciously by certain user intentions. Neither of these aspects should be ignored (see also Gelman & Hennig, 2017).

### 3.2 Data-analytic aims

Within the family of data-analytic aims, we draw a distinction between, on the one hand, aims related to the *subject* of the data analysis, that is to say, its goal and underlying principles, and, on the other hand, aims related to data-analytic *quality*. We will now successively discuss each of these two.

#### *3.2.1 Aims related to the subject of the data analysis*

Table 2 presents a series of taxonomic distinctions for the category of aims related to the subject of the data analysis. Within this category, we discern general and clustering-related aims.

Table 2
*Data-analytic aims related to subject of data analysis*

| **General aims** | |
|---|---|
| capture structural basis, process basis, or causal mechanisms underlying target variables | |
| detection of patterns in space or time | |
| identification of unusual/unexpected events, change points, … | |
| detection of patterns of association between variables | |
| capture functional relationship between predictor variable(s) and criterion variable(s) | |
| **Clustering-related aims/clustering principles** | |
| CLUSTERING PRINCIPLES BASED ON INTERNAL STRUCTURE | |
| *Perspective* | *Principles* |
| attributes/variables | |
| | within-cluster homogeneity with respect to defining characteristics |
| | between-cluster differences account for correlations between variables |
| (dis)similarities | |
| | large within-cluster similarities between cluster members |
| | large similarities between cluster members and cluster-representative object |
| | small between-cluster similarities |
| distribution in data space | |
| | cluster connectedness |
| | cluster shape |
| | each cluster concentrated on low dimensional manifold |
| | type of between-cluster separability |
| | cluster centroids in low dimensional subspace |
| | cluster area has high density |
| | cluster area is surrounded by near-zero density |
| CLUSTERING PRINCIPLES BASED ON RELATION WITH EXTERNAL VARIABLES | |
| within-cluster homogeneity | |
| between-cluster differences | |

General data-analytic aims are defined as aims that do not inherently involve a clustering. Table 2 contains a (non-exhaustive) list of five of such aims. Although there is no inherent link between these aims and clustering, most of them can be combined with the aim to induce clusters; optionally, this combined aim may then be addressed by means of clusterwise or mixture extensions of classical data-analytic methods. As examples one may think of: (1) mixtures of factor analyzers (Ghahramani and Hinton, 1997; McLachlan & Peel, 2000), which may address the wish to capture the structural basis underlying target variables in terms of factors, combined with a clustering, and (2) clusterwise regression (Hennig, 2000), to capture the functional relationship between predictor variables and a criterion, in conjunction with a clustering.

The second part of Table 2 lists clustering-related aims or clustering principles, based either on the internal cluster structure, or on the relation with one or more external variables; the internal-external distinction is based here on whether or not the variables are part of the information that will directly be used in the core data analysis. The taxonomic distinctions for the principles based on the internal structure go back, in part, to earlier divisions suggested by Van Mechelen et al. (1993) and Hennig (2016). They are organized according to the perspectives or angles from which one may look at the data that form the basis of the clustering and at the clusters themselves.

A first angle is that of features, attributes, or variables in the data. Clusters may then be required to be homogeneous with respect to cluster-defining characteristics, that is to say, (combinations of) feature-, attribute-, or variable-values. This may further lead to so-called monothetic clusters with singly necessary and jointly sufficient cluster-defining characteristics, or, stated differently, clusters with conjunctive definitions (e.g., Chavent, 1998). Alternatively, one may go for a clustering principle based on between-variable correlations rather than on (combinations of) variable-values. In particular, one may wish that between-cluster differences in variable-values account for correlations between variables, while within-cluster correlations between variables are zero (Rosch, 1978); this is exemplified by the local (i.e., within-cluster) independence assumption in latent class analysis.

A second angle is based on similarities or dissimilarities. This angle obviously requires a suitable specification of the respect(s) (dis)similarity has to capture. (Note that, from a formal viewpoint, (dis)similarities are defined in different ways in literature; here we use a least restrictive definition, with for similarities higher values referring to a higher similarity and for dissimilarities to a higher dissimilarity, without additional requirements with regard to range, symmetry, and triangle or other inequalities; "proximities" are further used as a collective term for both.) As for all clustering principles, the similarity-based principles may focus on the within- or the between-cluster structure.

A third angle is based on the distribution of the data points in a variable space. Most criteria within this angle are topological or geometrical in nature. Three criteria pertain to the form of the clusters: connectedness, assuming some particular shape such as a convex or ellipsoidal one, and concentration on a low dimensional subspace or manifold. The next two criteria pertain to the between-cluster structure: type of separability (e.g., linear), and the constraint that all cluster centroids lie in a low dimensional subspace (e.g., Markos et al., 2019; Vichi & Kiers, 2001). The final two criteria are density-related; the first of them involves a within-cluster aspect (viz., a cluster should correspond to a high density area), and the second a between-cluster aspect (viz., clusters should correspond to areas surrounded by a near-zero density area).

Next, we have principles based on the relation with external variables. Here, one may again focus on within- or between-cluster differences, and optionally also on the combination of both.

Importantly, for many data sets, different clustering principles may lead to considerably different clusterings. For example, there may be extended areas with a high density in data space, implying that clusters associated with high-density areas may host rather dissimilar members. Focusing on large within-cluster similarities will therefore lead to quite different clusterings than an emphasis on high-density areas. This makes it essential to get input from the user (which may require a considerable effort), without which the clustering problem is not well identified.

*3.2.2 Aims related to quality of the data analysis*

We now move to the category of aims related to the quality of the data analysis. Table 3 lists a series of relevant taxonomic distinctions, ordered as a function of the phase of the data analysis in the broad sense they focus on (input/data preprocessing, core data analysis, and output postprocessing); in addition, we also list three quality criteria that involve the relation between data-analytic input and output.

Table 3

*Data-analytic aims related to quality of data analysis*

| *Aspect of data analysis* | *Aim* |
|---|---|
| input preprocessing | |
| | include operations needed to address user concerns |
| | adequately factor in domain knowledge |
| core data analysis | |
| | include operations needed to address user concerns |
| | adequately factor in domain knowledge |
| | quality of algorithmic performance |
| output postprocessing | |
| | include operations needed to address user concerns |
| | adequately factor in domain knowledge |
| | parsimony (Ockham's razor) |
| relation between input and output | |
| | goodness of fit |
| | appropriate influence of different parts of the data on the output |
| | stability or replicability |

For all three phases of the data analysis, quality may mean that all operations needed to address the user concerns have been taken care of. This can be illustrated by means of the following examples: At the level of the data preprocessing, all variables relevant for the user concerns have been included; at the level of the core data analysis, a clustering criterion has been implemented that is relevant for these concerns; at the level of the output postprocessing, care has been taken that the output contains all required information to address the user concerns (e.g., that it includes results of postprocessing analyses to investigate the relation between the output of the core data analysis to external variables of interest to the user).

In a similar vein, one may wish the three phases of the data analysis to adequately factor in all relevant domain, background, or subject-matter knowledge. This knowledge may include a fully-fledged substantive-theoretical account of how the data came about, all available information on it, and all consequences implied by it. We first illustrate at the level of the data preprocessing with an example from metabolomics (i.e., the large-scale study of the concentrations in cells, tissues, or biofluids of metabolites, that is, small molecules that are substrates or products of the metabolism of the organism under study). Substantive theories in this domain imply that the concentration of a set of metabolites (variables) in a set of tissues (objects), measured in terms of ion counts in mass spectrometry, is subject to so-called instrumental response factors (i.e., metabolite-specific relations between number of ion counts and metabolite concentration); this implies that concentration measurements are not comparable across metabolites. In this case, an appropriate preprocessing to restore comparability may involve a transformation based on so-called calibration information (Van Mechelen & Smilde, 2011). A second illustration pertains to the phase of the core data analysis and concerns fluorescence spectroscopy data. One may wish the mathematical structure of the model underlying the core data analysis to adequately represent the so-called “conceptual model” underlying the data generation (i.e., the substantive framework that allows for a semantic interpretation of the terms in the mathematical model: see Harshman & Lundy, 1984); this appears to be the case for a PARAFAC analysis (Van Mechelen & Smilde, 2011). Thirdly, in the phase of the output postprocessing, substantive theories may, for instance, imply constraints on the choice of the number of clusters in the postprocessing stage.

An additional criterion that may be considered at the level of the core data analysis is algorithmic performance. The latter may include aspects such as computational speed, scalability, convergence speed, and safety measures against algorithms getting trapped in inferior local optima.

With regard to the relation between the input and the output of the core data analysis, a standard quality criterion is that of goodness of fit (which may be relevant for nonstochastic as well as stochastic models); in the output postprocessing phase, this may often be considered in conjunction with the criterion of parsimony, because of an obvious tradeoff relation. A second quality criterion at the level of the input-output relation pertains to an “appropriate” (relative) influence or impact of different parts of the data on the output; different possible parts at this point include particular objects or experimental units (e.g., “outliers”), particular types of variables (e.g., continuous vs. categorical in mixed variable setups, see also Hennig & Liao, 2013), and particular types of data blocks (pertaining to particular sources, or particular types of information). As a third criterion with regard to the input-output relation, one may consider various kinds of stability (Leisch, 2016; Ullmann et al., 2022) and replicability across different studies (Masoero et al., 2023); stability is a precursor of replicability, and replicability is most topical, given the replicability crisis in several domains of science (e.g., Romero, 2019).

### 3.3 Data

Table 4 presents a list of taxonomic questions on the data, ordered along five different aspects. In all cases, the starting point is a single data entry, that is to say, the filling of a single cell in the data array. As running examples one may think of the expression level of some gene in some tissue, and of the degree of similarity between two brands as judged by some consumer.

A first aspect is the **structure of the data**, based on old but seminal distinctions drawn by Tucker (1964) and Carroll & Arabie (1980). Within this context, a first question is which the different sets involved in a single data entry are. In the gene expression example, these are the set of the genes and the set of the tissues, and in the brand similarity example the set of the brands and the set of the consumers. If for a certain set, the elements of that set show up in different roles in a data entry (as in network communication data, where the same persons show up in the roles of sender and receiver) that will subsequently be associated in the data analysis with different structural characteristics (viz., parameters), we consider each role to be associated with a different set (Carroll & Arabie, 1998).

Table 4
*Taxonomic questions on data.*

| Aspect | Taxonomic question |
|---|---|
| structure of data | |
| | What are the different sets of elements that are involved in a single data entry? |
| | What kind of combination of the sets is involved in a single data entry? |
| | Are data entries in principle available for all possible combinations of elements? |
| | Are some data modes structured in terms of prespecified characteristics? |
| | Do the data comprise information of different kinds on the same set of entities? |
| Values of data entries | |
| | Are the data entries single values? |
| | In case of single numeric values: do all numeric relations/operations make sense? |
| | Which data entries can be meaningfully compared across the data array? |
| Data availability | |
| | Are all data available at once? |
| Prior membership info | |
| | Is prior information on class membership available? |
| | If prior information is available, is it fully available? |
| Data-related challenges | |
| | Are the data large / ill-conditioned? |
| | Are the data sparse? |
| | Do the data contain outliers? |
| | Are there (many) missing values? |
| | Are there any problems with data quality? |

Subsequently, a second question is which kind of combination of the sets is involved in a single data entry. In our examples, the expression level pertains to an ordered pair of a gene and a tissue, and the brand similarity rating to an ordered triplet of two brands and a consumer. (Optionally, the answers to the first two questions can be formalized by conceiving the data as a mapping from a Cartesian product of a number of sets to some value set, which may also be referred to as a data block or data tensor.) The number of distinct sets involved in a data entry (or in the Cartesian product) is called the number of *modes*, and the total number of sets involved in an entry (or in the Cartesian product) is called the number of *ways*. As such, the gene expression data can be considered two-way two-mode, and the brand similarity data three-way two- mode. Standard data types for cluster analysis are two-way two-mode object by variable data $d_{ov}$, $o = 1, ..., O; v = 1, ..., V$, and two-way one-mode object by object proximities $d_{oo'}$, $o, o' = 1, ..., O$ (Figure 2), with "object" being defined as the (physical or nonphysical) unit of primary interest in a study.

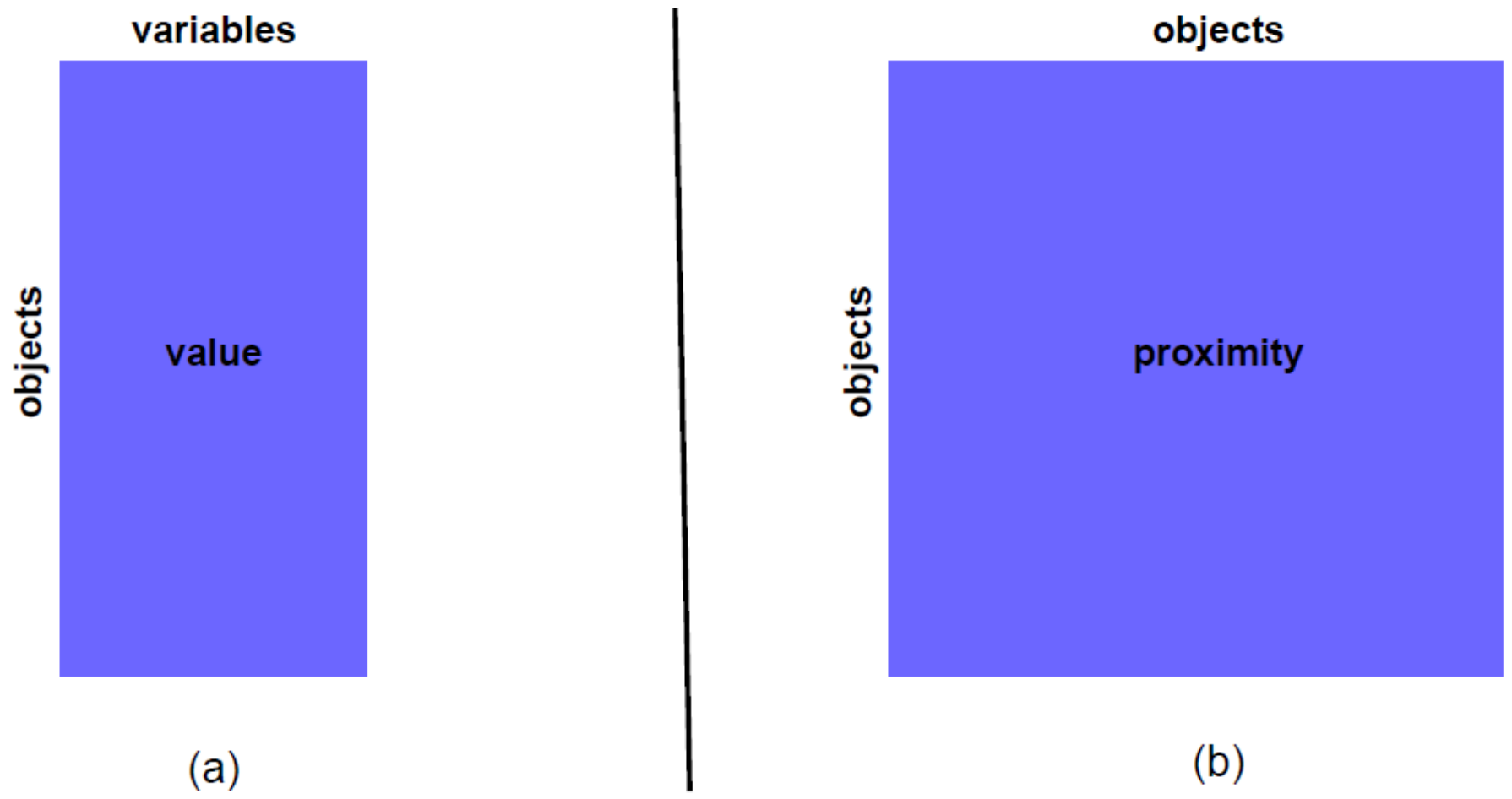

*Figure 2.* Standard data structures for cluster analysis: (a) two-way two-mode object by variable data $d_{ov}$, (b) two-way one-mode object by object proximity data $d_{oo'}$.

Third, one may wonder whether data entries are, in principle, available for all combinations of elements of the different data ways (i.e., for all ordered *n*-tuples of their Cartesian product); “in principle” here refers to the fact that at this point we leave incidental missingness aside. If this is the case, the data are called fully crossed. Alternatively, the data may be subject to some kind of structural missingness. Examples include: (1) between-subject experiments in which each participant is being assigned to one of the experimental conditions only, and where the values of the dependent variable(s) of interest for the other conditions are therefore structurally missing, and (2) nested (multiset) data in, for instance, cross-cultural studies, where values on variables are being recorded for participants from different cultures, and where, for a given participant, values on these variables are structurally missing for cultures other than the culture of the participant in question.

The fourth question is whether one or more data modes are structured in terms of some available prespecified characteristic. Basic examples of such characteristics are location in time or space, as in longitudinal or spatial data (e.g., Caiado et al., 2016; Handl et al., 2016; Hitchcock & Greenwood, 2016); a more complex example of a characteristic arises if the variables that constitute a variable mode are hypothesized to be manifest indicators of a few latent constructs (Everitt et al., 2011).

The fifth structural question is whether the data comprise information of different kinds on the same set of entities, which may stem from, for instance, different sources or measurement platforms. If this is the case, indeed, so-called coupled or multiblock data are at stake, as studied in problems of data integration or data fusion (Van Mechelen & Smilde, 2010). Coupled or multiblock data are data that consist of two or more data blocks, with the different data blocks being connected to one another via common modes of pairs of blocks. To illustrate, Figure 3 presents four hypothetical examples of coupled data. For instance, Panel (a) of this figure represents so-called node-attributed network data as studied in node-attributed network clustering (e.g., Chunaev, 2020) as a simple example of multiblock data. It consists of a two-way actor by actor matrix (which contains information on the intensity of the interaction between each pair of actors) coupled with a two-way two-mode actor by actor attribute matrix (which contains information on the characteristics of the actors, i.e., the attributes of the nodes in the social network). In this example, the two data blocks have the actor mode in common (which therefore acts as the linking mode between the two data blocks).

Regarding the **values of the data entries**, one may first of all wonder whether or not they are single values. As an alternative, they may be sets, intervals, or distributions. These are treated in the literature on “symbolic data” (Brito, 2014, 2016) and “uncertain data” (Aggarwal, 2014b). As an example of symbolic data, one may think of two-way two-mode store by variable data, in which the variable “Customer age” takes as values intervals or histograms. From their part, uncertain data refer to uncertainty with regard to an individual data entry or all data entries that pertain to an element of a data mode (e.g., due to imprecise instruments) that may be captured by means of a univariate or multivariate probability distribution.

In case of data entries that are single values, a further distinction can be drawn in terms of their assumed measurement level (Stevens, 1946; Velleman & Wilkinson, 1997). The primary difference here is that between numeric values for which all numeric relations and operations are assumed to make sense, and numeric or nonnumeric values for which this is not the case. Beyond this, further distinctions between measurement levels have been made in order to delimit sensible mathematical operations to be carried out on the measurements (see Mosteller & Tukey, 1977, for a typology that goes beyond the seminal one of Stevens, 1946). In many situations, the measurement level issue is subtle and can be controversial. A key issue is whether the measurements are the result of a mapping from the world to the numerical system that is well-defined within a measurement-theoretical framework (with the relationship between numbers preserving the observed empirical relationships between objects), or are a merely pragmatic representation (Hand, 2004; Roberts, 1979). In any case, the nature of the measurements has implications on how data can sensibly be processed, and the user is well advised to take this into account.

Somewhat related to the measurement level issue, one may also wonder which data entries can be meaningfully compared across the data table (Van Mechelen & Smilde, 2011); the latter also goes with the label of (un)conditionality. Unconditional data are data all entries of which can be meaningfully compared, whereas row-conditional data, column-conditional data, etc. are data for which comparisons are only meaningful within rows, columns, etc. As a first example, one may think of two-way two-mode object by variable data where the different variables are expressed in different measurement units; this implies that only comparisons within variables (columns) are meaningful, implying that the data are variable- or column-conditional. Secondly, we may revisit the example of the tissue-by-metabolite concentration data from the previous section; in this case all variables (metabolites) can be expressed in the same measurement unit (e.g., number of ion counts in mass spectrometry); nevertheless, prior to any preprocessing, comparisons across variables are in this case not meaningful because of so-called instrumental response factors, which imply that the data are once again column-conditional in nature. A third example, also from metabolomics, is even more subtle; it pertains to longitudinal or time-resolved two-way two-mode biofluid-by-time metabolite intensity date (for a single metabolite only). Suppose now that some of the biofluids pertain to blood, whereas other ones pertain to urine of the same individual. In that case, comparisons within a given time point across biofluids are not meaningful (which implies biofluid- or row-conditionality); the reason for this is that the meaning of a time point is not the same across biofluids, as metabolites usually appear earlier in the blood. Stated differently, in this case, comparisons across biofluids are hampered by the fact that the biological rather than the physical time is relevant (Van Mechelen & Smilde, 2011).

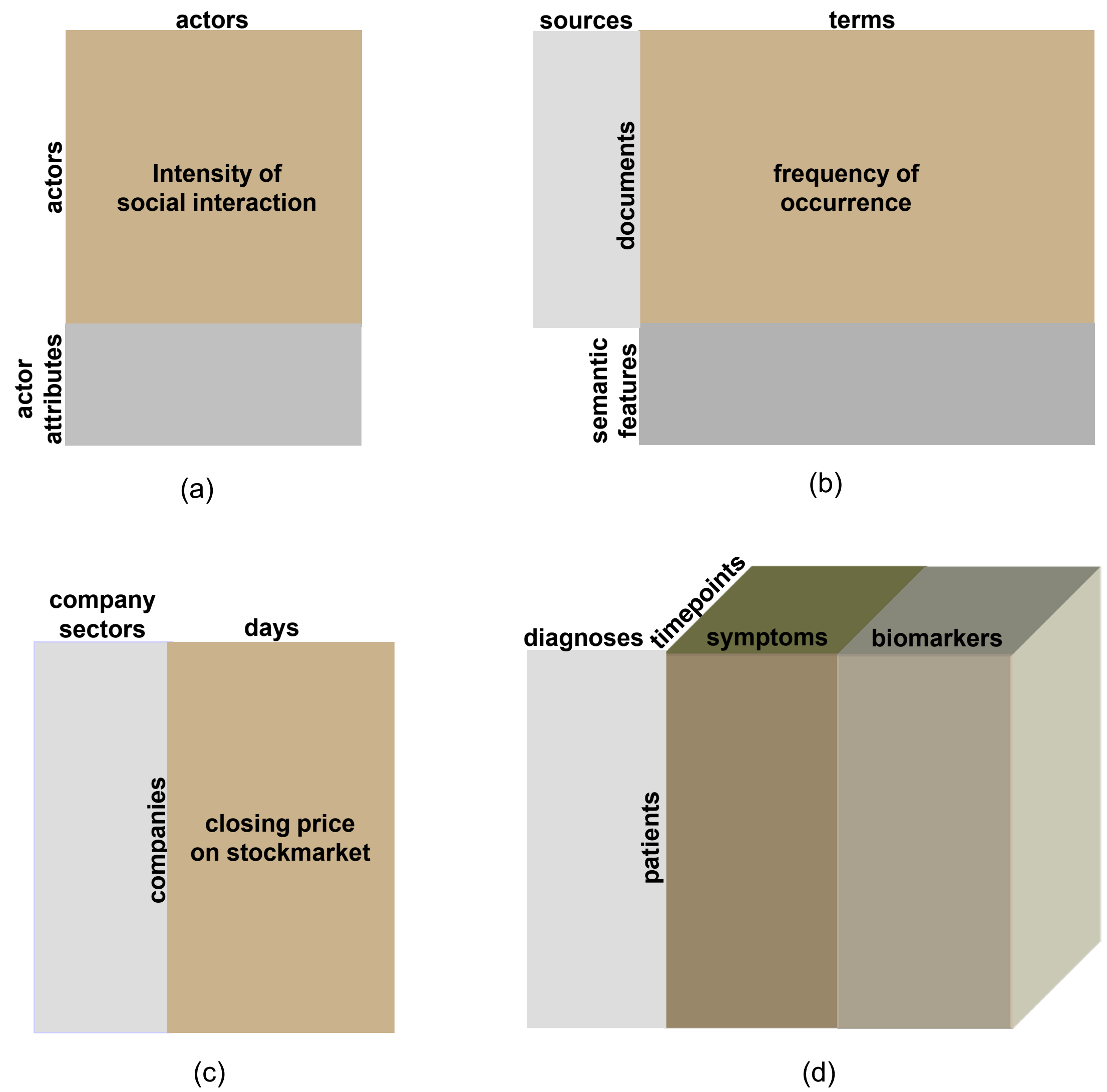

*Figure 3*. Four hypothetical examples of coupled data: (a) two coupled data blocks as studied in node-attributed network clustering (e.g., Chunaev, 2020); (b) three coupled data blocks as studied in document clustering; (c) and (d): (two resp. three) coupled data blocks as studied in time series clustering.

Regarding **data availability**, one may wonder whether all data entries are available at once. If yes, static data are at play; if not, we have to deal with dynamic data or data streams (Aggarwal, 2014a). As examples of the latter, one may think of e-commerce data or data on social network activity.

**Prior information on class membership** may or may not be available, and, if available, the information may be partial or full. This corresponds to the classical distinction between unsupervised problems (no information on class membership available, as in a prototypical clustering setting), semi-supervised problems (partial information available: see, e.g., Bair, 2013; Jain et al., 2016), and supervised problems (full information available, as in a prototypical classification setting). Incidentally, in (semi-)supervised settings, (unsupervised) cluster analysis may still be of interest, as one may, for instance, wish to know how well a data-driven clustering corresponds to the fully or partially given grouping.

Fifth and finally, **data-related challenges** may be present, in addition to data issues already mentioned before such as lack of comparability and data that are not fully available at once. Challenges include: (1) large data (i.e., data that involve a large number of variables and/or objects) and ill-conditioned data (i.e., data in which the variables considerably outnumber the objects) which imply considerable computational, technical and inferential complications, including giving rise to potentially less meaningful clusters that are in an absolute sense large in size but in a relative sense very tiny, (2) sparse data, that is, data that contain many zeroes (as, e.g., data resulting from many problems on text mining), (3) data that include outliers (as already referred to under quality of the data-analysis/input-output relation), (4) data that include missing values, with missingness here being incidental (in contrast to the structural missingness discussed before), and (5) problems with data quality in terms of high amounts of noise or high numbers of distortions. All possible combinations of these challenges may show up as well.

### 3.4 Input preprocessing

We now turn to a descriptive account of the data analysis in the broad sense and the many choices that show up in it. We start with the preprocessing of the input of the core data analysis, that is to say, of the data. Table 5 lists a number of relevant choices for two aspects of it: basis and level.

Table 5
*Choices in input preprocessing.*

| Aspect | Choice | |
|---|---|---|
| Basis | | |
| | internal data | |
| | external data | |
| Level | | |
| | involvement of data parts | |
| | | retaining vs. dropping |
| | | Weighting |
| | changes within initial data structure | |
| | | explicitly or implicitly replacing values in (parts of) the data |
| | | addition of parts to the data |
| | modification of initial data structure | |

The **basis** of the preprocessing may be internal or external data, meaning data that will versus will not be directly involved in the core data analysis. As an example of preprocessing based on internal data, one may think of standardizing variables. As an example of preprocessing based on external data, we revert to the earlier example of tissue-by-metabolite concentration data that were prone to metabolite-specific instrumental response factors, the influence of which could be wiped out by means of a preprocessing based on external calibration information. Incidentally, the distinction between preprocessing based on internal versus on external data is not without consequences: In case of an internal basis, preprocessing could be taken care of in an "automated" way, whereas preprocessing with an external basis (usually rooted in domain knowledge) can only proceed in a user-based, nonautomated manner.

Preprocessing further can take place at three different **levels**. A first of these pertains to the involvement of specific parts of the data, such as particular variables, objects, data cells, and so on. Choice alternatives here are to drop (vs. retain) them and, somewhat more subtly, to give them a lower weight for the purpose of the subsequent core data analysis. As an example, one may think of removing or downweighting objects that take outlying or extreme values on one or more variables.

A second possible level of preprocessing is that of changes within the initial data structure without modifying the type of that structure. A first choice alternative at this point is to replace values in particular parts of the data (rows, columns, data blocks, …) by other values (typically on the basis of part-specific transformations). As an example, one may think of variable-specific standardization. Note that, most often, such transformations are taken care of in an explicit way. Sometimes, however, one may opt for transforming the data by nonlinearly mapping them to a virtual high dimensional Euclidean feature space (in which one may hope clusters to be linearly separable, unlike in the original data space), without making this mapping explicit; although in such cases the explicit form of the transformation in question is unknown, Euclidean distances in the feature space are nevertheless calculable thanks to the so-called “kernel trick” (Xu & Wunsch, 2009). As another example, linear and nonlinear dimension reduction and embedding methods such as principal component analysis and (*t*-distributed) stochastic neighbor embedding (van der Maaten & Hinton, 2008) are sometimes applied to preprocess data before clustering (although it cannot be taken for granted that the information in the original data that is lost in this way is irrelevant for clustering). A second choice alternative involves an addition of parts to the data; examples of this type include the addition of new variables that are sums or logical combinations of old variables, and single or multiple imputations of missing values.

A third, and most extensive level of preprocessing involves a modification of the type of the initial data structure. As examples at this level, one may think of conversions of (two-way two-mode) object by variable data into (two-way one-mode) object by object proximities (for which one could rely on all possible kinds of proximity measures: see, e.g., Gower & Legendre, 1986), and of matricizing three-way three-mode data, that is to say, converting them to a two-way two-mode matrix by concatenating two of the original data modes.

We conclude this section with two overall remarks: Firstly, some issues may be taken care of either during the preprocessing or during the core data analysis. As an example, one may think of variable-specific transformations that cannot only be taken care of during the preprocessing, but also during the actual data analysis (viz., in so-called optimal scaling methods: see, e.g., van Buuren & Heiser, 1989). As another example, one may think of performing dimension reduction and clustering sequentially (also referred to as tandem analysis) versus simultaneously (e.g., Labiod & Nadif, 2021). Secondly, within the stage of preprocessing, some issues may be dealt with in different ways. As an example, missing values can be taken care of by means of removing objects or variables, or by means of (single or multiple) imputations.

### 3.5 Core data analysis

In this section we will specifically focus on the cluster analysis part of the core data analysis. The taxonomic distinctions we will draw (some of which have already been briefly mentioned in Van Mechelen et al., 2023) will nevertheless also be relevant for the clustering part of data-analytic methods that serve, in addition to finding or constructing groups, other user concerns and data-analytic aims. A discussion and taxonomic organization of data-analytic methods without a clustering part, however, falls outside the scope of the present paper.

Cluster analysis may be looked at both from the angle of the mathematical structure or the model it relies on (be it a probability model or a nonstochastic model), and from the angle of the algorithm or computational procedure behind it. These two angles are, on the one hand, to be distinguished from one another; on the other hand they are also closely interrelated (with not infrequently confusion arising from the fact that the same name is used to denote both a cluster-analytic model and an associated cluster-analytic algorithm). Some disciplines primarily describe cluster analysis from the first angle and other ones from the second angle. Obviously, both angles are important. Nevertheless, when drawing taxonomic distinctions with regard to the cluster analysis part of the data analysis, we will primarily do so from the viewpoint of the mathematical structure underlying the cluster analysis, and in particular from a set-theoretical perspective.

To draw these distinctions, we start from the fundamental observation that clusters have two sides: The first of these is the cluster membership; the second one is the cluster meaning, the set of cluster characteristics, the cluster definition, and the cluster membership rule. In cognitive science, the first usually goes with the term “category” and the second with the term “concept” (Smith & Medin, 1981; Van Mechelen et al., 1993). Relatedly, within a long philosophical tradition (Leibniz, 1764), the first is associated with the term *extension*, and the second with the term *intension*. Mathematically speaking, the two sides correspond to the two ways in which a set can be specified: either by enumerating its elements, or by specifying its defining rule. One may note that all of the above is most straightforward within an implicit context of object by attribute data, where the concept or intensional side of a cluster is easily recognizable. The distinction, however, is also relevant within all other possible data contexts, including that of object by object proximities. To be sure, whereas all clustering procedures yield extensional output, this is not always the case for intensional output: The retrieval of the latter does not seldom require additional analyses based on internal or external information in the postprocessing stage. Furthermore, in our account below, we will extend the concept of cluster intension, by also subsuming cluster-specific representative objects, features of the within-cluster structures, and within-cluster models under it.

Table 6 lists a number of critical questions for a taxonomic organization of data-analytic methods that take care of the cluster analysis part of the core data analysis. The first part of the table pertains to questions on the clusters' extension, and the second to questions on intensional aspects.

Table 6
*Questions on core data analysis.*

| Questions on clusters' extension (cluster membership) |
|---|
| What is the nature of the elements of the clusters? |
| Is cluster membership crisp or fuzzy? |
| Does the method aim at one or a few clusters or at a clustering to be assessed as a whole? |
| Are all elements to be clustered? |
| Are overlapping clusters possible? |
| Does the method aim at a single clustering or at multiple clusterings? |
| Do (intrinsic or extrinsic) extensional constraints apply? |
|     Are they constraints on what may be admissible clusters? |
|     Are they constraints on what may be admissible clusterings? |
| **Questions on clusters' intension (cluster meaning)** |
| Does each cluster go with a pattern of attribute-values, a centroid, prototype, or representative object(s)? |
| Does each cluster go with a cluster membership rule in terms of characteristics or attributes? |
| Does each cluster go with a cluster-specific weight that represents feature importance or salience? |
| Does each cluster go with a cluster-specific model? |
| Do intensional constraints apply? |

*3.5.1 Extensional level*

At the extensional (cluster membership) level, a first question is "What is the nature of the elements of the clusters?" Elements can be objects, variables, or some other type of entities involved in the data (for example, sources). In case of nested data, the elements to be clustered can be situated on any hierarchical level. In case of data that involve two or more types of entities, such as object by variable or source by object by variable data, the elements to be clustered can also be ordered *n*-tuples such as ordered pairs (of, e.g., an object and a variable) in bi-, two-mode, or co-clustering (see, further, e.g., Bock, 1968; Hartigan, 1975; Madeira & Oliveira, 2004; Van Mechelen et al., 2004; Vichi, 2016). One may note that it is possible that a clustering method simultaneously yields multiple sets of clusters with different types of elements. For instance, many biclustering methods for object by attribute data simultaneously yield biclusters of ordered object-attribute pairs, in addition to clusters of objects and clusters of attributes implied by the biclusters (e.g., Van Mechelen et al., 2004).

A second question is whether crisp or fuzzy cluster membership is at stake. The standard form of cluster membership is crisp in nature, with elements not belonging versus belonging to a cluster under study, formalized by membership values of 0 versus 1. Alternatively, graded forms of membership can be considered, with membership values varying from 0 to 1; optionally, this could further be formalized in terms of some fuzzy set-theoretic framework (e.g., Bezdek et al., 2005). At this point clustering methods that yield as output cluster membership probabilities need special attention, as the probabilities may be part of two clearly different scenarios: Firstly, they may be secondary output under the form of posterior cluster membership probabilities that may be derived from primary output under the form of estimates of model parameters; in that case the probabilities in question represent our inherently limited knowledge about crisp memberships at the level of an underlying latent model. Secondly, the probabilities may be model parameters that can be interpreted as fuzzy/graded membership values, indeed.

Third, one may wonder whether the cluster analysis aims at one cluster or a few clusters versus at a clustering that is to be assessed as a whole. Certain cluster analysis methods aim at retrieving a single cluster that in itself is optimal in some sense (e.g., Ben-Dor et al., 2002; Calonder et al., 2006); as a variant, some among them subsequently look for the second (third, fourth, etc.) best cluster, possibly after putting aside its previously detected predecessor(s). A far more common alternative is to look for a collection of clusters, that is, a clustering, which is to be assessed as a whole in some respect; in the latter case, the number of clusters in the set can be either prespecified or can be chosen during or after the actual cluster analysis.

If one is looking for a clustering that comprises two or more clusters, three follow-up questions may further be asked. A first of these is whether in the cluster analysis method under study all elements are to be clustered. If this is the case, the clustering constitutes what is called an exhaustive clustering or a cover; if not, the clustering may be considered to include a residual category of unclassified elements. A second question is whether the cluster analysis method may provide a clustering in which overlap between clusters is possible. Overlap is to be read at this point as *extensional* overlap, that is to say, a single element belonging simultaneously to two or more clusters (e.g., Shepard & Arabie, 1979). The combination of a cover with absence of extensional overlap yields a partition, assuming that all clusters are nonempty. A partition implies that the rows of the element by cluster membership matrix, both in the crisp and in the fuzzy case, sum to 1 (which is also called row stochasticity). A third question is whether the cluster analysis method yields a single clustering or multiple alternative clusterings. As to the latter, one may think of simultaneously inducing multiple different clusterings that all take comparable values on some objective function of interest (e.g., Bailey, 2014; Jain et al., 2008). That being said, the output of some cluster analysis methods may be considered both as a single clustering and as a set of alternative clusterings. As an example, one may think of hierarchical cluster analysis methods, the output of which could be looked at as a single hierarchical clustering or as a series of alternative (nested) partitions. In such cases, we go for the (subjective) convention to consider the output as a single clustering only.

A final group of extensional questions is whether (hard or soft) extensional constraints apply (with “hard” vs. “soft” referring to whether the constraints have to be met fully vs. to some degree only). Such constraints may, firstly, pertain to what might be *admissible clusters*, and may be based on intrinsic or extrinsic information. At this point, intrinsic (in contrast to extrinsic) information is defined as information that exclusively stems from the clustering. Examples of intrinsic information-based constraints on cluster admissibility include a prespecified minimum threshold on cluster size (cardinality), the requirement that clusters are maximal or not expandable, and that biclusters of object by attribute data are constrained to rectangles or Cartesian products of a row and a column cluster. Examples of extrinsic information-based constraints on cluster admissibility include must-links and cannot-links (i.e., constraints on whether certain pairs of elements must or cannot show up in one and the same cluster), and contiguity constraints (in, e.g., time or space) on elements that belong to one and the same cluster (Basu et al., 2009). Secondly, extensional constraints (based again on intrinsic or extrinsic information) may apply to what might be *admissible clusterings*. Within this category, examples of intrinsic information-based constraints may include: (1) constraints on the number of clusters, (2) balancing constraints on the difference in sizes of different clusters in a clustering, (3) nestedness constraints on overlapping clusterings, meaning that overlap is limited to clusters in a subset-superset relation (possibly in addition to the requirement that the clustering also contains all singletons as well as the full set of elements), (4) pyramidal constraints on overlapping clusterings (Bandelt & Dress, 1989; Diday, 1986), and (5) the constraint that multiple alternative clusterings should be mutually as different as possible. Finally, examples of extrinsic information-based constraints on clustering admissibility may include that the final clustering is inspired by prior knowledge (e.g., via an informative prior in a Bayesian approach: Paganin et al., 2021), and that alternative clusterings are as similar as possible to (or as different as possible from) a prespecified, given clustering, or something in-between these two (e.g., Qi & Davidson, 2009).

*3.5.2 Intensional level*

Next, we move to the cluster intension. Questions at this level primarily pertain to the availability and nature of intensional information. As such, a first question is whether each cluster goes with a pattern of attribute-values (either in a crisp or in a graded sense), with a single centroid or representative object, or with multiple representative objects. One may observe that the notion of a single centroid could be associated with the so-called prototype view on categories in cognitive science, according to which the meaning of a category is stored in the mind through an idealized representation of the category center in terms of a list of weighted attributes; likewise, the notion of multiple representative objects could be associated with the so-called exemplar view that goes with a representation in terms of multiple sub-prototypes (Hampton, 2015).

Secondly, one may wonder whether each cluster goes with a cluster membership rule that is specified in terms of cluster characteristics or attributes (i.e., a so-called declarative membership rule), and, if yes, what would be the form of this rule. Alternative possibilities at this point include conjunctive rules (in terms of a set of singly necessary and jointly sufficient features), and sums of the weights of the features that are present in the object under study (which possibly have to exceed some prespecified threshold).

A third question is whether each cluster goes with a cluster-specific weight that represents the importance or salience of that cluster when conceived as a latent feature, with cluster membership representing presence of the feature in question. This is the case in so-called latent binary feature models (Bandelt & Dress, 1989; Van Mechelen & Storms, 1995).

Fourth, one may wonder whether each cluster goes with a cluster-specific (nonstochastic or stochastic) model, with cluster-specificity meaning that at least part of the model parameters or of the structural model characteristics would be allowed to vary across clusters. At this point, any family of simple or complex models can qualify. Examples of models include simple distributions (as in, e.g., multivariate normal mixtures: see, e.g., Bouveyron et al., 2019), regression models (as in clusterwise regression), component or factor analysis models (e.g., McLachlan & Peel, 2000), and hidden Markov models (e.g., Coviello et al., 2014). For the special case of biclustering, the issue of modeling the within-bicluster structure deserves some proper attention: At the level of nonstochastic modeling, one possible option is that all entries within a bicluster assume the same reconstructed data value, which may be prespecified (e.g., 1 in bicluster models for 0/1 data) or may be to be estimated. As an alternative one may consider nonconstant biclusters with particular within-bicluster patterns (in terms of, e.g., a bicluster-specific set of row or column main or multiplicative effects). At a stochastic level, all kinds of block mixture models may be considered (e.g., Govaert & Nadif, 2008; Riverain et al., 2023).

Fifth, one may wonder whether (soft or hard) intensional constraints apply. Examples include constraints on the within-cluster structure (e.g., in terms of admissible types of within-cluster covariance matrices, or of the same within-cluster covariance matrix for all clusters). Otherwise, in Bayesian clustering (Wade, 2023), intensional as well as extensional constraints can be imposed through the specification of suitable prior distributions.

### 3.6 Output postprocessing

The final step in (one cycle of) the data-analytic process is the output postprocessing. Table 7 lists a number of relevant specifications to taxonomically structure this step, organized in terms of three aspects: target, basis, and tools.

Table 7
*Aspects of output postprocessing.*

| Aspect | Possible specification |
|---|---|
| Target | |
| | Model selection |
| | Model evaluation |
| | Output interpretation |
| Basis | |
| | Internal information vs. external information |
| | Transformed data-analytic output |
| Tools | |
| | Visualization/graphical tools |
| | Numerical tools |

In general, the **target** may pertain to model selection, model evaluation, and output interpretation; note that we use the term "model" here in a general sense, referring to any formal structure that acts as the footing of the data analysis. The **basis** for specifying or addressing the target can be purely internal (i.e., information that is involved in the core data analysis) or may also include external information. Moreover, addressing the target may or may not involve transformations of the data-analytic output; examples of such transformations include converting posterior cluster membership probabilities into zero/one memberships, back-transforming preprocessed variables to their raw counterparts, and, if the output comprises multiple clusterings, a search for a consensus or ensemble clustering (e.g., Affeldt et al., 2020; Ghosh & Acharya, 2016). Third, **tools** may include various kinds of visualization techniques (e.g., making use of suitable projections) as well as a broad range of numerical techniques based on indices, statistical tests, and various kinds of (multivariate) data-analytic procedures.

More specifically, with regard to *model selection*, a first and ubiquitous question pertains to determining the number of clusters $K$ (which also includes the selection of a partition in a sequence of nested partitions). An important sub-question is whether the number of clusters $K$ equals 1, which comes down to the question whether any clustering structure is present. The number of clusters issue can be addressed on the basis of internal or external information, while making use of a broad range of tools, including visualization methods, formal tests, information criteria, and all possible kinds of heuristics or “stopping rules”. For any particular tool, a point of interest, though, is whether it also allows to conclude that $K$=1. Next to determining the number of clusters, many other model selection aspects may show up. In particular, this may be the case for: (1) hypotheses on parameters or structural characteristics of cluster-specific models (such as the shape of within-cluster covariance matrices, or the numbers of factors in mixtures of factor analyzers), and (2) whether external information-based constraints such as must- and cannot-links are not at odds with the data.

*Model evaluation* (the clustering part of which is often referred to as cluster validation) can be based on internal or external information:

- Evaluation based on internal information may first of all pertain to goodness of fit between the data and the model output, which may be looked at both from a global or omnibus viewpoint and from all possible kinds of specific viewpoints. Second, stability may be investigated in many respects (Ullmann et al., 2022), by means of influence analyses, investigations of stability with regard to analyzing parts of the data only, or making use of resampling techniques (Leisch, 2016), and investigations of stability with regard to alternative possible options in the data preprocessing and core data analysis (in terms of sensitivity and multiverse analyses: see Steegen et al., 2016). Third, examination of one or more internal validity indices is further also subsumed under internal information based model evaluation, with different such indices typically targeting different clustering-related aims (Halkidi et al., 2016). Fourth, visualization techniques can be used to assess to what extent the found clusters correspond to separated, clearly distinguished patterns in the data (e.g., Hennig, 2004).
- Evaluation based on external information comprises, first of all, investigating the relation between the clustering resulting from the core data analysis and all external variables that are of interest to the user. For this purpose, in addition to visualization methods, various data-analytic techniques could be invoked, such as analysis of variance, discriminant analysis, and ordinary or multinomial logistic regressions to predict cluster membership on the basis of external covariates. Alternatively, one may also wish to compare the obtained clustering with external clusterings (e.g., clusterings resulting from previous research) making use of suitable indices (e.g., the Adjusted Rand Index or Normalized Mutual Information in case of partitions, see further Meila, 2016, and the correlation between cophenetic matrices in case of hierarchical clusterings, see Sokal & Rohlf, 1962). Otherwise, such a comparison could also be taken care of in a prospective way, by examining the extent to which an obtained cluster structure can be replicated in a future study.

*Output interpretation*, too, can be addressed on the basis of both internal and external information. Moreover, it can also be dealt with by means of many kinds of graphical and numerical tools.

As a final remark, a number of postprocessing aspects can be taken care of during both the postprocessing and the core data analysis. To subsume such aspects under output postprocessing can therefore be considered a somewhat arbitrary convention. As examples one may think of: (a) several aspects of model selection, including the choice of the number of clusters, that in several approaches are dealt with during the core data analysis –incidentally, deciding whether the data are heterogeneous enough for clustering ($K$=1 vs. $K$>1) is even framed in the literature on “clusterability” (Adolfsson et al., 2019) as something to be addressed *before* the core data analysis–; (b) relating a clustering result to external variables of interest to the user during the core data analysis (e.g., Kim et al., 2016), which, otherwise, implies that the variables in question become, by definition, internal information; (c) interpreting clustering solutions, which can not only be taken care of post hoc (as, e.g., in common explainable AI or XAI approaches) but also during the core data analysis (as in, e.g., so-called intrinsically explainable AI approaches: Peng et al., 2022).

## 4. ILLUSTRATIVE EXAMPLES ON USE OF THE OUTLINE MAP IN PRACTICE

In this section we will showcase how the proposed outline map could be used in practice. We will do so for three types of illustrations: We will show how the map could act as an aid in (1) passing through the full cluster-analytic process, (2) a sharp characterization of a single clustering method, and (3) a sharp characterization of the relation between two clustering methods.

### 4.1 Aid in passing through the full cluster-analytic process

Smith and Ehlers (2020) ran a clinical psychological study on grief trajectories after the bereavement of a significant other. Below we will typify the different steps in their study based on the six constituents of our outline map as summarized in Figure 1. (The result of this typification is further also graphically represented in supplementary Figure 1Sup in Appendix B.)

*User concerns*. The authors started by stating that they aim “to investigate to what extent a comprehensive set of (…) cognitive factors (…) predicts the course of grief severity in the first 12 to 18 months after a loss” (Smith & Ehlers, 2020, p. 94). For this purpose, they first wanted to identify “subgroups or classes of individuals who show similar longitudinal patterns on (…) grief severity”, as they expected that "this approach allows a more nuanced perspective on the way grief changes over time (…) than currently afforded by the diagnostic system [that tells apart] clinical vs. nonclinical grief" (Smith & Ehlers, 2020, p. 94). Subsequently, they wanted to examine whether cognitive measures predict class membership. Within our framework, one may distinguish within these user concerns a general concern of prediction in conjunction with a grouping-related concern of finding groups. Clearly, the grouping-related concern figures among the original concerns. It further appears to act as a means to achieve a prediction goal; because of this we could label it as a secondary concern.

*Data-analytic aims*. Related to the subject of the data analysis, one may distinguish a clustering-related aim based on the internal clustering principle of within-cluster homogeneity with respect to defining characteristics and a general aim of capturing the functional relationship between predictor variables and a cluster membership criterion (with the latter to be created during the data analysis). Related to the quality of the data analysis, three further aims may further be distinguished: (1) the input preprocessing and output postprocessing should include all operations needed to address the user concerns, and all relevant domain knowledge should be factored in, with operations including the calculation of scale scores during the preprocessing, and domain knowledge being involved in the choice of variables during the postprocessing, (2) quality of algorithmic performance in terms of safety measures against the occurrence of local optima, and (3) goodness of fit.

*Data.* In terms of structure, the data consist of two blocks that comprise information of two different kinds on the same set of participants. The first data block contains grief severity ratings on 11 items of the Prolonged Grief Disorder Inventory from 275 participants who recently experienced a bereavement, at 3 time points (at time of recruitment, 6 months later, and 12 months later); the structure of this block is therefore three-way three-mode data, with the modes being items, participants and time points; there is no structural missingness, and one mode (viz., that of the time points) is structured in terms of a prespecified characteristic (time order). The second data block contains predictor information in terms of ratings on 5 background variables and items of scales to capture 4 cognitive and 4 interpersonal measures by 275 participants at a single time point (viz., time of recruitment); the structure of this block is therefore two-way two-mode data, with the modes being variables and participants; there is no structural missingness and one mode (viz. that of the variables) is structured in terms of background information on which item belongs to which scale. The data entries are single numerical values for which all numeric relations/operations are assumed to make sense. An assumption of variable conditionality has been made; this implies, for instance, that in Block 1 the scores on an item at time of recruitment are assumed to be comparable across all participants, and, hence, that the psychological meaning of time of recruitment is assumed to be the same for all participants (which is not obvious, especially because the time interval between bereavement and recruitment was not constant across participants). Finally, all data were available at once, prior information on class membership was not available, and no major data-related challenges have been reported.

*Input preprocessing*. The preprocessing includes a summation of the items of the Prolonged Grief Disorder Inventory, of the items of each of the 4 cognitive measures and of the items of each of the 4 interpersonal measures, leading to 9 scale scores; this preprocessing was based on external information on which items go together and comes down to an addition of parts of the data; upon replacing each set of items by the corresponding scale, Block 1 has been reduced to a two-way two-mode matrix (modification of initial data structure). The three time moments have further been treated as equidistant and as comparable across participants, which has been translated into a recoding of time into values of 0, 1, 2. The preprocessed two-way two-mode grief severity data block can then be denoted by $Y_{it}$, $i = 1, ..., 275$; $t = 0,1,2$ , with $Y$ denoting grief severity, $i$ the index of the individual participant and $t$ that of time.

*Core data analysis*. This analysis was putatively based on the following linear growth mixture model:

$$\left(Y_{it}, t = 0,1,2\right) \sim N\left(\left(b_{0i} + b_{1i}t, t = 0,1,2\right); \sigma^2 \mathbf{I}\right) \;,\; \left(b_{0i}, b_{1i}\right) \sim \sum_{k=1}^{K} \lambda_k N\left(\beta_{0k}, \beta_{1k}; \Sigma\right),$$

with $N\left(\boldsymbol{\mu}\Sigma\right)$ denoting a multivariate normal distribution with mean vector $\boldsymbol{\mu}$ and variance-covariance matrix $\Sigma$ , $\left(b_{0i}, b_{1i}\right)$ a random intercept and slope that are assumed to be distributed according to a mixture of $K$ bivariate normals with common covariance matrix Σ, with $\lambda_k$ $\left(k = 1, ..., K\right)$ denoting the prior probability of mixture component (i.e., cluster) $k$, and $\mathbf{I}$ denoting the identity matrix. As the algorithm used by the authors to estimate this model could end up in a local optimum, they dealt with this by means of a multistart procedure. From an extensional viewpoint, the model can be characterized as follows: The elements of the clusters are participants, cluster membership is crisp, the method aims at a comprehensive clustering, all elements are to be clustered without extensional overlap (hence a partition is at stake), and the method aims at a single clustering without extensional constraints. From an intensional viewpoint, each cluster goes with a cluster-specific linear grief severity profile, defined by the parameter pair $(\beta_{0k}, \beta_{1k})$; there are no explicit declarative cluster membership rules nor cluster-specific weights that represent latent feature importance or salience (the prior probabilities or mixing parameters $\lambda_k$ do not qualify as such); each cluster goes with a cluster-specific linear growth model; there are no intensional constraints.

*Output postprocessing*. With regard to model selection, the final number of clusters was selected on the basis of a combination of several goodness-of-fit measures; the linearity constraint on the growth models was examined informally only via inspection of goodness-of-fit indices for a one-cluster linear growth model. Model evaluation based on internal information was taken care of in terms of inspection of goodness of fit, yet considered from a relative viewpoint only, in terms of a comparison of models with 1 through 5 clusters. An operation in the form of a transformation of the data-analytic output was carried out, by converting the posterior cluster membership probabilities into zero/one memberships. A model evaluation based on external information then involved multinomial logistic regressions to predict the zero/one cluster memberships on the basis of the background variables, and the scale scores for the cognitive and interpersonal measures.

From the above, one may safely conclude that our proposed outline map allows, indeed, to characterize in an in-depth way the different aspects of a real-life data-analytic process, and to profoundly understand its architecture and the many choices made in it. To be sure, we set up our analysis in a purely descriptive way. Alternatively, however, a similar analysis could also be set up from an evaluative perspective. Within such an approach, one could ask questions such as: “Are the choices that have been made suitable?” (As examples within the context of the present study, one may think of the choices to go for a clustering and to assume variable conditionality.) What would be possible alternatives? Which alternatives are to be preferred?

**4.2 Aid in sharp characterization of a single clustering method**
We will illustrate this second type of application with three brief case descriptions.

*4.2.1 Global kernel k-means clustering*
This method, proposed by Tzortzis and Likas (2008), can be summarized as *k*-means, combined with a nonlinear mapping of the data into a high dimensional feature space (that can remain implicit thanks to the so-called “kernel trick”), and a structured multistart based on the *k*-1 centroids of the final solution of (*k*-1)-means, plus, as $k^{th}$ centroid, the vector of variable values of each individual object. We can characterize the method within the proposed outline map as follows:

*Data-analytic aims.* Related to the subject of the data analysis, one may note a clustering-related aim based on the internal clustering principle of within-cluster homogeneity with respect to defining characteristics, albeit in the high dimensional feature space, which cannot be readily back-translated to the original data space; in addition, one might also note an implicit aim based on the distribution in the high dimensional feature space, that is to say, linear separability. Related to the quality of the data analysis, an aim with regard to the core data analysis is at stake, that is to say, quality of algorithmic performance in terms of safety measures against the occurrence of local optima; this is addressed by means of the structured multistart. An additional aim at the level of algorithmic performance is to reduce, insofar possible, the high computational complexity and cost.

*Data.* This method operates on one-block, two-way two-mode object by variable data, without structural missingness, and with unstructured data modes. The data entries are supposed to be single numeric values, on which all numeric operations make sense, and that are unconditional (i.e., comparable across the full data matrix). All data are expected to be available at once, prior information on class membership is expected not to be available, and there are further no specific data-related challenges.

*Input preprocessing.* The method involves a preprocessing based on internal data via an implicit transformation, that is to say, an implicit mapping of the variables to a higher dimensional feature space (kernel trick).

*Core data analysis*. From an extensional viewpoint, the method produces a single crisp partition of the objects without extensional constraints. From an intensional viewpoint, cluster centroids are situated in the high dimensional feature space and typically remain implicit; there are further neither cluster-specific membership rules, weights or models, nor intensional constraints.

The above illustrates how a characterization of a method on the basis of the proposed outline map may allow to quickly highlight the key characteristics of that method. Moreover, by doing so one may easily move beyond the name of the method: In “global kernel *k*-means”, “global” in itself is not a very transparent label, and “kernel” may give rise to confusion with clustering methods based on kernel density estimators.

#### *4.2.2 Fuzzy DBSCAN*

Kriegel and Pfeifle (2005) proposed an adaptation of DBSCAN for uncertain object by object distance data, with uncertainty being captured by a probability distribution for each data entry that is given as a part of the data. From its part, given a radius *r* and a minimum cardinality *m*, DBSCAN builds a cluster starting from a “core object”, that is, an object for which the hypersphere with as center that object and with radius *r* contains at least *m* objects; the initial cluster associated with that core object then is the set of objects in the hypersphere in question. Subsequently, the method recursively extends the current cluster with all radius *r* hyperspheres centered at a core object in that current cluster. Fuzzy DBSCAN extends DBSCAN on the basis of probabilistic extensions of the concepts of “being a core object” and “belonging to a radius *r* hypersphere”. We can characterize the method within our proposed outline map as follows:

*Data-analytic aims*. These are internal structure based clustering-related aims with regard to the subject of the data analysis; in particular, from the perspective of the distribution of the objects in the data space, clusters are conceived as high density areas surrounded by near-zero density areas.

*Data*. The method applies to one-block two-way one-mode object by object distance data, without structural missingness, and with unstructured data modes. The data entries are probability distributions expressing data entry uncertainty. The values are assumed to be unconditional, that is, comparable across the whole data matrix. All data should be available at once. There is no prior information on class membership, and there are no specific data-related challenges.

*Input preprocessing*. The method does not involve or require a particular type of preprocessing, except if it would have to be applied to two-way two-mode object by attribute data; in that case, a conversion to two-way one-mode object by object distances would be required, possibly after a suitable rescaling of the variables to make them comparable.

*Core data analysis*. From an extensional viewpoint, the data analysis relies on a single crisp clustering of objects, with not all objects to be clustered as the analysis leaves room for “noise objects” that do not belong to any cluster; extensional overlap is further not allowed, and there are no extensional constraints. From an intensional viewpoint, each cluster goes with a set of core objects; there are no cluster-specific membership rules, weights, or models, and no intensional constraints.

As illustrated again above, via the characterization on the basis of the outline map, misunderstandings with regard to key characteristics of a method can be avoided; as an example, the “fuzzy” in Fuzzy DBSCAN appeared not to refer to cluster membership but to data uncertainty. Also, certain aspects are sharpened; as an example, the “density-based” in DBSCAN appears to include also the density-based clustering principle of separation (i.e., clusters are to be surrounded by a near-zero density area), unlike in some mode-seeking procedures based on kernel density estimators.

#### *4.2.3 The plaid model*

The plaid model, as originally proposed by Lazzeroni and Owen (2002), and fine-tuned by Turner et al. (2005), produces a biclustering of object by attribute data $d_{ij}$ (denoting the value of object *i* for attribute *j*) under the form of a set of rectangular, overlapping biclusters; the biclustering further also includes a background bicluster consisting of the Cartesian product of the full object set and the full attribute set. Each bicluster further goes with a heterogeneous set of values that follow a “plaid pattern” defined by bicluster-specific object and attribute main effects, with the pattern of the background cluster acting as a kind of offset term. The reconstructed data entry $\hat{d}_{ij}$ then equals the sum of the bicluster-specific (*i*,*j*) values of all biclusters to which (*i*,*j*) belongs. A characterization of the method according to the outline map yields the following:

*Data-analytic aim*. The data analysis targets the general aim of capturing the structural mechanism underlying the data, with biclusters playing a key role in this mechanism. Specifically, within the context of gene by condition expression level data, the biclusters may be interpreted as combinations of co-regulated genes and conditions under which these genes are up- or down-regulated (Lazzeroni & Owen, 2002).

*Data*. The method presupposes one block of two-way two-mode object by attribute data, without structural missingness, and with unstructured data modes. The data entries should be single numeric values, on which all numeric operations make sense, and which can be assumed to be unconditional. All data should be available at once. Optionally, prior information on class membership may be partially available. Within a systems biological context, a typical data-related challenge is that the data may be high dimensional (which will be dealt with in the estimation of the plaid model via a computational procedure that involves simple parameter updates only).

*Core data analysis*. From an extensional viewpoint, the analysis produces a single crisp biclustering of ordered pairs of an object and an attribute; extensional overlap between the biclusters is allowed; extensional constraints further include that each bicluster should be the Cartesian product of a row and a column cluster, and that the biclustering contains a background bicluster consisting of the Cartesian product of the full object set and the full attribute set (the latter implies that the biclustering is a cover). From an intensional viewpoint, each bicluster goes with a bicluster-specific within-bicluster pattern of values defined by bicluster-specific object and attribute main effects; there are no cluster-specific membership rules or weights, or intensional constraints.

As illustrated above, the characterization on the basis of the outline map may facilitate a fast and relatively deeper understanding of a method, especially also for readers who are less familiar with methods of a similar type.

**4.3 Aid in sharp characterization of the relation between two clustering methods**

We illustrate by means of the relation between the latent stochastic blockmodel (Nowicki & Snijders, 2001) and the mixed membership stochastic blockmodel (Airoldi et al., 2008). Stochastic blockmodels are limited here to models for directed graph data $Y_{i\,i'} \in \{0,1\}$, $i \neq i'$, where *i, i'* denote the vertices/objects to be clustered, and with assumed independence between $Y_{i\,i'}$ and $Y_{i'\,i}$. The latent stochastic blockmodel can be conceived as a mixture model., Each mixture component implies a latent partition of the objects into *K* classes. Within each mixture component, independence of the $Y_{i\,i'}$-variables is assumed, and

$$Y_{i\,i'}|_{\text{object } i \,\in\, \text{class } k \text{ and object } i' \,\in\, \text{class } k'} \sim \text{Bernoulli}\left(\theta_{k\,k'}\right),$$

with $\theta_{k\,k'}$ denoting the probability of a directional link from an element of class *k* to an element of class *k'*. The mixed membership stochastic blockmodel subsequently has been introduced as a generalization of the latent stochastic blockmodel that aims at removing the ”limitation that each object can only belong to one cluster” (Airoldi et al., 2008, p. 1982).

Within the framework of our proposed outline map the relation between the two methods can be characterized as follows:

*Data-analytic aim*. Both methods target the general aim of capturing the structural basis underlying directed graph data (in terms of clusters of objects, with the objects within a given cluster having a similar interaction pattern with the other objects within the same and within the other clusters).

*Data.* Both methods operate on one-block, two-way one-mode object by object data, without structural missingness, and with unstructured data modes; note that, in the two methods, corresponding row and column objects are associated with the same parameters, which implies that we should consider the sets of the row and the column objects as a single set (one mode). The data entries are single 0/1 values, which are assumed to be unconditional. All data are available at once; there is no prior information on class membership, and no specific data-related challenges are at stake.

*Core data analysis*. From an extensional viewpoint, for the latent stochastic blockmodel the analysis relies on a single *crisp* partition of the objects without extensional constraints; in contrast, for the mixed membership stochastic blockmodel, it relies on a single *fuzzy* partition of the objects without extensional constraints. From an intensional viewpoint, in both methods each cluster goes with a “centroid”, that is to say, a specific pattern of interactions with the other clusters, as represented by the pair of vectors $\left(\boldsymbol{\theta}_{k.}, \boldsymbol{\theta}_{.k}\right)$ corresponding to the $k^{\text{th}}$ row and the $k^{\text{th}}$ column of the $\Theta$-matrix ; there are further no cluster-specific membership rules, weights or models, and no intensional constraints.

*Postprocessing.* The latent stochastic blockmodel usually goes with a postprocessing step to produce crisp partition class memberships.

Making use of the above characterization, we can now further specify the core data generation mechanism in each of the two methods. The latent stochastic blockmodel starts from a latent crisp partition of the objects into *K* classes; this may be formalized by means of a membership vector of length $K$ for object $i$, $\mathbf{z}_i$, with $z_{ik} \in \{0, 1\}$, $\sum_{k=1}^{K} z_{ik} = 1$, and assuming $\mathbf{z}_i \overset{iid}{\sim} \text{Multinomial}(\lambda_1, ..., \lambda_K)$; subsequently, the data entries $Y_{i\,i'}$ may be generated, with $Y_{i\,i'} \sim \text{Bernoulli}(\mathbf{z}_i\, \Theta\, \mathbf{z}_{i'}^{'})$, and $\Theta$ denoting a $K \times K$ matrix with entries $\theta_{k\,k'}$ that represent the probability that a class $k$ object is linked to a class $k$' object. In contrast, the mixed membership stochastic blockmodel starts from a latent fuzzy partitioning of the objects into $K$ classes; this may be formalized with the help of a membership vector of length $K$ for object $i$, $\mathbf{m}_i$, with $m_{ik} \in [0, 1]$, $\sum_{k=1}^{K} m_{ik} = 1$, and assuming $\mathbf{m}_i \overset{iid}{\sim} \text{Dirichlet}(\alpha_1, ..., \alpha_K)$, where $(\alpha_1, ..., \alpha_K)$ denotes a vector of hyperparameters to be estimated; subsequently, the data entries $Y_{i\,i'}$ may be generated by firstly drawing for each ordered pair $(i, i')$ crisp partition membership vectors $\mathbf{z}_i$ and $\mathbf{z}_{i'}$, with $\mathbf{z}_i \sim \text{Multinomial}(m_{i1}, ..., m_{iK})$, and subsequently drawing $Y_{i\,i'}$ from $Y_{i\,i'} \sim \text{Bernoulli}(\mathbf{z}_i\, \Theta\, \mathbf{z}_{i'}^{'})$; again, as in the latent stochastic blockmodel, $\Theta$ denotes a $K \times K$ matrix of probabilities $\theta_{k\,k'}$ that a class $k$ object is linked to a class $k$' object.

All the above shows clearly that the key difference between the two methods under study resides in the latent partition of the objects at the start of the data generation, which is crisp in the latent stochastic blockmodel and fuzzy in its mixed membership counterpart. Note also that in this regard the claim that the mixed membership stochastic model removes the limitation that each object can only belong to one cluster is potentially misleading, as the method relies on a *partition*, albeit a fuzzy one. Note that models with probabilistic (fuzzy) cluster memberships $m_{ik} \in [0, 1]$ *without* the partitioning constraint $\sum_{k=1}^{K} m_{ik} = 1$ do exist as well (see, e.g., Maris et al., 1996; Meulders et al., 2001).

## 5. DISCUSSION AND CONCLUSION

### 5.1 Added value of the proposed approach

We started this paper from the finding that the clustering domain suffers from a major accessibility problem as well as the allied problem that the domain is rife with division. The domain, indeed, comprises many pretty isolated islands, in spite of the fact that over more than 50 years calls for unification and building bridges have been launched (e.g., Arabie, 1982) based on the belief that such bridges would lead to most welcome synergies.

To address these problems, we made use of the traditional toolkit of the cluster analyst: clustering, building taxonomies, and drawing sets of taxonomic distinctions. For this purpose, we relied in part on existing concepts and distinctions, which we here brought together within a single, systematic, and unifying framework.

The chosen approach has three distinctive characteristics: Firstly, the clustering domain has been examined and characterized from *multiple perspectives*. Although a full crossing of all possible specifications within each of these perspectives might not be possible, it is nevertheless possible to look at the cluster-analytic process (resp. at a single clustering method or at two or more of such methods) from several of these perspectives, and to examine and characterize them accordingly. Otherwise, one may note that quite a few handbooks of cluster analysis include chapters based on one value for one taxonomic distinction within one perspective only (e.g., “clustering methods for uncertain data”, “semi-supervised clustering”, “kernel-based clustering”). Secondly, the chosen approach was *first principles in nature*, in that it did not start from existing, traditional subdivisions, but from concepts and distinctions that are as fundamental as possible, and that have been defined and delineated as sharply as possible. Thirdly, the approach chosen was *disconnected from specific substantive domain contents*. To be sure, this does not mean that substantive concerns, concepts, research questions, and theories, as well as substantive domain knowledge are not relevant or important in the cluster-analytic process; quite the contrary, as evidenced by many references to them throughout the paper.

We believe that our proposed approach has a threefold added value: Firstly, with regard to the *data-analytic process*, it provides a kind of “hold” in terms of a structure, an analysis of the different steps in the process, and lists of possible choice alternatives as well as a kind of “source of unrest” under the form of a strong incentive to thoroughly think through the whole. In spite of attempts to streamline the clustering workflow (Esnault et al., 2023; Kamoshida & Ishikawa, 2020), which might go with obvious advantages on the level of communication and reproducibility, such a thorough thinking through the data-analytic process will never be possible in a fully automated way. Indeed, as explained before, clustering problems are cursed with a major lack of identifiability, and every solution for this will inevitably require decisions from the user, not only in order to remove the identifiability obstacle, but also in order to fully take the user's aims into account. In this respect, relevant questions may include: “What do you want as a user or as a data-analyst?”, “What do you care about?”; “Which data do you need for this?”; “How should you process these data?”, “Which path would you better take?” (in the knowledge that not all, but usually many paths are legitimate, and with the subsequent challenges of taking/combining several legitimate paths, and of examining whether they converge in terms of conclusions). Secondly, with regard to the *characterization of single methods*, as illustrated above, the proposed approach may facilitate to capture core aspects of a given method, and to arrive at a deeper understanding of it, averse from labels and terms that could be used in a misleading way (such as “fuzzy” and “overlap”). Such a characterization may considerably facilitate a better match between user concerns and methods. Thirdly, the proposed approach may also facilitate a deeper understanding of *relations between methods*, including methods that originate from different research domains. Hopefully, the latter may also contribute to answering the calls referred to above for building bridges between the different (sub)disciplines that are involved in the study and development of cluster-analytic methods.

**5.2 Gaps and future directions**

An obvious limitation of the present paper is that we only gave proof-of-concept illustrations of how our proposed approach could be used in practice (see Section 4). As such, the application of the proposed conceptual framework to the vast range of cluster-analytic methods implies a major challenge for future research. Anyway, the latter also goes with a standing invitation to characterize existing as well as newly to be developed methods on the basis of the taxonomic distinctions we have proposed here, in view of a more in-depth understanding of these methods as well as their interrelations. A similar invitation pertains to a thoughtful planning and evaluation of applications of cluster analysis in data-analytic practice. Furthermore, to facilitate the use of our proposed approach for these different purposes, we subsumed all of the distinctions drawn in the present paper in a six-part checklist (see Appendix C), the six parts of which correspond to the six ellipses as represented in Figure 1. In line with the illustrative examples of Section 4, to characterize/ evaluate/analyze/guide a cluster-analytic process, all six parts of the checklist have to be filled out, whereas to characterize/analyze/compare clustering methods (including introducing new methods), this has to be done for parts 2 through 5.

We close with three final remarks. Firstly, given the magnitude of the challenge addressed in the paper, our proposed framework and sets of taxonomic distinctions can only constitute an onset of a solution. A further elaboration of this solution will undoubtedly be needed. As examples, one may think of the addition of taxonomic distinctions within each of the outlined perspectives, and of the addition of more specifications/options within each of the taxonomic distinctions. Secondly, we wish that our proposal will stimulate further research in the direction of formalizing and measuring how well specific aims of clustering (as in the checklist in Appendix C) are fulfilled. Indeed, most existing criteria for evaluating the quality of a clustering have been presented with an overly general ambition of measuring overall quality, without differentiating between different aspects of potential interest, as done in our paper; there is much need for more "targeted" criteria. In general, we cannot but send a standing invitation to all research communities involved in the clustering domain to critically examine and discuss the proposals launched here, and to further elaborate them. Thirdly, the focus of the present paper was on cluster-analytic methods. Nevertheless, a number of concepts and distinctions put forward in the paper may also have relevance for other subdomains of statistics, data analysis, data science, and computer science. As examples, one may think of the backbone scheme of the data-analytic process, and of the taxonomic distinctions with regard to data.

**Funding information**

The research reported in this paper was supported in part by the Research Foundation - Flanders (grant K802822N to Iven Van Mechelen), by the Research Council KU Leuven (SAB/22/050), and by the Department of Statistical Sciences of the University of Bologna (visiting grant 1393/2022 to Iven Van Mechelen).

**Acknowledgements**
The first author gratefully acknowledges the hospitality of the University of Bologna and the University of Groningen during two study visits that provided ample opportunity for the research at the basis of this paper. The authors thank Bedilu Alamirie Ejigu, Hendrik Blockeel, Hans Bock, David Hand, and Boris Mirkin for most helpful comments and suggestions on an earlier version of this paper.

## APPENDIX A: GLOSSARY

**admissible clusters**: clusters that meet imposed (intensional or extensional) constraints
**algorithmic performance**: quality of algorithmic functioning in terms of computational speed, scalability, convergence speed, and safety measures against the occurrence of local optima
**alternative clustering analysis**: cluster analysis method that simultaneously induces multiple different clusterings that all take comparable values on some objective function of interest
**balancing constraint**: constraint that different clusters in a clustering are approximately the same size
**biclusters**: clusters of ordered object-variable pairs (in case of two-way two-mode object by variable data)
**cannot-links**: constraints on whether certain pairs of elements cannot show up in one and the same cluster
**Cartesian product (of sets $S_1, \ldots, S_n$)**: set of all ordered $n$-tuples ($s_1, \ldots, s_n$), with $s_1 \in S_1, \ldots, s_n \in S_n$
**cluster-analytic algorithm**: computational procedure that yields a cluster or a clustering
**cluster-analytic model**: nonstochastic or stochastic mathematical structure at the basis of a cluster analysis
**cluster ensembles**: see consensus clustering
**clustering**: collection of clusters that is to be assessed as a whole
**cluster validation:** evaluation of cluster-analytic model based on internal or external information
**co-clusters**: see biclusters
**column-conditional data**: data for which comparisons of entries are only meaningful within columns
**consensus clustering**: combining multiple clusterings of the same set of objects into a consolidated summary clustering
**constructing groups:** data-analytic concern pertaining to inducing groups from data that are required for practical or pragmatic reasons, whereas an interpretation in terms of an unobserved meaningful underlying distinction is not relevant
**contiguity constraint**: constraint that elements that belong to one and the same cluster are contiguous (in, e.g., time or space)
**core data analysis**: phase of data analysis between input preprocessing and output postprocessing
**cover**: clustering in which all elements of the set under study are to be clustered
**coupled data**: data that comprise information of different kinds on the same set of entities
**crisp clusters**: clusters with elements that either belong or do not belong to it, formalized by membership values of 0 versus 1
**data-analytic aims**: aims related to the subject of the data analysis (i.e., its goals and underlying principles) and aims related to data-analytic quality
**data block**: part of data set that can be formalized as a mapping from a subset of a Cartesian product of sets to some value set
**data entry**: filling of a single cell in the data array
**data streams**: data all entries of which are not available at once
**data tensor**: see data block
**declarative cluster membership rule**: cluster membership rule that is specified in terms of cluster characteristics or attributes
**dissimilarity**: measure of closeness of two objects with higher values referring to a lower closeness
**dynamic data**: see data streams
**evaluation**: user concern to evaluate something according to some evaluative standard
**exemplar view**: view on categories in cognitive science, according to which the meaning of a category is stored in the mind in terms of multiple category members or sub-prototypes
**explanation**: user concern pertaining to uncovering the causes underlying the phenomenon of interest in terms of immediate precursors or of factors at the basis of the phenomenon’s genesis, as well as mechanisms and processes underlying it
**exploration**: user concern to discover unexpected elements or patterns in the data
**extension**: cluster membership; in cognitive science also referred to as “category”
**external data/variables**: data/variables that are not part of the information that is directly used in the core data analysis
**extrinsic information (-based constraints on what might be admissible clusters/clusterings)**: (constraints in terms of) information that does not exclusively stem from the clustering
**finding groups:** data-analytic concern pertaining to inducing groups from data that can be interpreted in terms of a meaningful underlying distinction **fully crossed data**: data for which entries are, in principle, available for all combinations of elements of the different data ways
**fuzzy clusters**: clusters with graded forms of membership varying from 0 to 1 that constitute primary output of the clustering procedure (unlike secondary output under the form of posterior cluster membership probabilities that represent inherently limited knowledge about crisp memberships at the level of an underlying latent model)
**fuzzy partition**: row-stochastic fuzzy clustering
**grouping-related concern**: concern of finding or constructing groups on the basis of data
**hard constraints (on clusters)**: (extensional or intensional) constraints that have to be fully met
**identification**: user concern pertaining to questions on the identity of the objects of interest
**ill-conditioned data**: data in which the variables considerably outnumber the objects
**information compression**: user concern pertaining to reducing an abundance of information, either for oneself or for communication to others
**input preprocessing**: phase of data analysis pertaining to the preparation of the data
**intension**: set of cluster characteristics (attributes, features), cluster definition, and declarative cluster membership rule in terms of cluster characteristics; in cognitive science also referred to as "concept"
**internal data/variables**: data/variables that are part of the information that is directly used in the core data analysis
**intrinsic information (-based constraints on what might be admissible clusters/clusterings)**: (constraints in terms of) information that exclusively stems from the clustering
**general data-analytic aims**: aims that do not inherently involve a clustering
**large data**: data that involve a large number of variables and/or objects
**linking mode**: common mode of two data blocks in coupled data
**modes (number of)**: (number of) distinct sets involved in a single data entry
**multiblock data**: see coupled data
**must-links**: constraints on whether certain pairs of elements must show up in one and the same cluster
**nested**: a collection of sets is called nested if in any nondisjoint pair of sets of the collection one of the sets is a subset of the other
**node-attributed network data**: two-way object x object interaction intensity data coupled with two-way object x attribute data, with the objects to be represented as nodes in a network
**object**: (physical or nonphysical) unit of primary interest in a study
**output postprocessing**: phase of data analysis pertaining to operations on the data-analytic results in view of presenting and interpreting them
**overlapping clustering**: clustering with clusters that are allowed to be nondisjoint (i.e., to display extensional overlap)
**partition**: cover consisting of nonempty, mutually disjoint clusters
**prediction**: user concern pertaining to any type of forecasting
**preparation of action**: user concern pertaining to the fact that the user intends to undertake some action and wants to optimally prepare for this
**primary grouping-related concern**: grouping-related user concern that figures among the initial concerns, either as the only concern, or in combination with one or more general concerns the importance of which is lower than or equal to that of the grouping-related concern
**prototype view**: view on categories in cognitive science according to which the meaning of a category is stored in the mind through an idealized representation of the category center in terms of a list of weighted attributes
**proximity**: collective term for similarity and dissimilarity
**pyramid**: clustering with associated (reconstructed) distance matrix $\Delta$ that satisfies the pyramid constraint, meaning that there exists an order $\prec$ on the set of elements which is such that for each triplet of elements $e_1, e_2, e_3$ with $e_1 \prec e_2 \prec e_3$: $\delta(e_1,e_3) \geq \text{Max}(\delta(e_1,e_2), \delta(e_2,e_3))$
**row-conditional data**: data for which comparisons of entries are only meaningful within rows
**row-stochasticity**: crisp or fuzzy clustering for which each row of the element by cluster membership matrix sums to 1
**secondary grouping-related concern**: grouping-related user concern that figures among the initial concerns in combination with one or more general concerns the importance of which is strictly greater than that of the grouping-related concern **semi-supervised problems**: learning problems in which prior criterion information (e.g., class memberships) is partially available
**similarity**: measure of closeness of two objects with higher values referring to a higher closeness
**soft constraints (on clusters)**: (extensional or intensional) constraints that have to be met to some degree
**sparse data**: data that contain many zeroes
**static data**: data all entries of which are available at once
**structural missingness:** data that do not include entries for all combinations of elements of the different data ways because of nonincidental, structural reasons
**supervised problems**: learning problems in which prior criterion information (e.g., class memberships) is fully available
**symbolic data**: data the entries of which are not single values but sets, intervals, or distributions
**two-mode clusters**: see biclusters
**uncertain data**: data that go with uncertainty about data entries, which may be captured by means of a univariate or multivariate probability distribution
**unconditional data**: data all entries of which can be meaningfully compared
**unsupervised problems**: learning problems in which prior criterion information (e.g., class memberships) is not available at all
**user**: person with a substantive domain interest at the basis of the data-analytic process
**user concerns**: what is important to the user, what the user aims at, needs, wishes, or hopes to get ultimately from the data analysis
**ways (number of)**: (total number of) sets involved in a single data entry

## APPENDIX B: SUPPLEMENTARY FIGURE

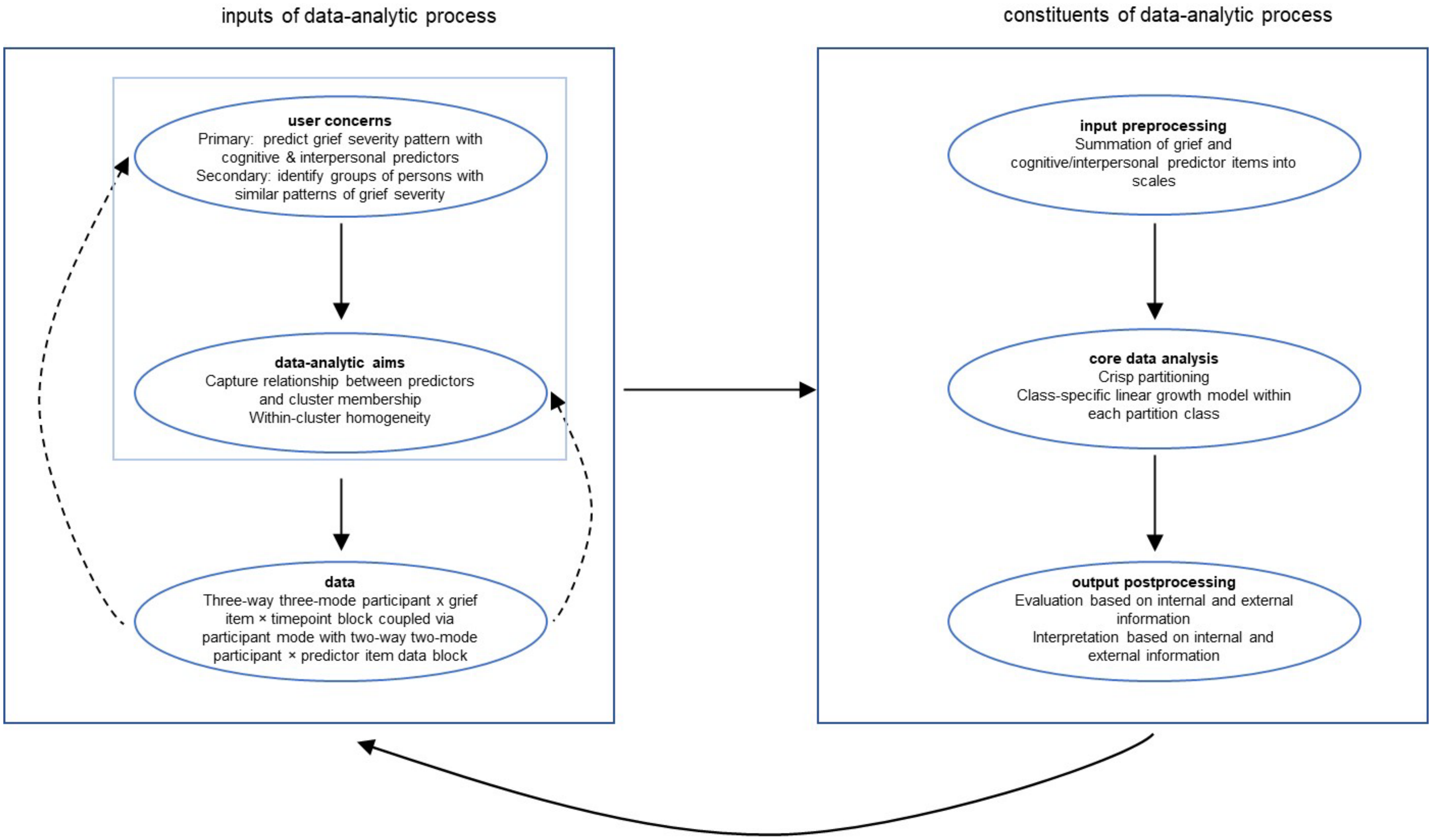

*Figure 1Sup.* Example set of fillings for the six ellipses of Figure 1 based on a conceptual analysis of the Smith and Ehlers (2020) study as presented in Section 4.1.

**APPENDIX C: CHECKLIST**

To characterize/evaluate/analyze/guide a cluster-analytic process, all six parts have to be filled out.
To characterize/analyze/compare clustering methods, parts 2 through 5 have to be filled out.

**Part 1: User concerns**

| *Aspect/Sub-aspect* | *Question* | *Check* | *Specification* |
|---|---|---|---|
| General concerns | | | |
| Content | | | |
| | Is explanation at stake? | ☐ | ............................. |
| | Is identification at stake? | ☐ | ............................. |
| | Is prediction at stake? | ☐ | ............................. |
| | Is preparation of action at stake? | ☐ | ............................. |
| | Is information compression at stake? | ☐ | ............................. |
| | Is exploration at stake? | ☐ | ............................. |
| | Is evaluation at stake? | ☐ | ............................. |
| Grouping-related concerns | | | |
| Nature | | | |
| | Is finding groups at stake? | ☐ | ............................. |
| | Is constructing groups at stake? | ☐ | ............................. |
| Relative importance | | | |
| | Is grouping concern primary original concern? | ☐ | |
| | Is grouping concern secondary original concern? | ☐ | |
| | Is grouping concern not among original concerns? | ☐ | |

**Part 2: Data-analytic aims**

| *Aspect/Sub-aspect* | *Question* | *Check* | *Specification* |
|---|---|---|---|
| Related to subject of data analysis | | | |
| General | | | |
| | Capture structural basis, process basis, or causal mechanisms? | ☐ | ………………………… |
| | Detection of patterns in space or time? | ☐ | ………………………… |
| | Identification of unusual/unexpected events, change points, …? | ☐ | ………………………… |
| | Detection of patterns of association between variables? | ☐ | ………………………… |
| | Capture functional relationship between predictor and criterion variable(s)? | ☐ | ………………………… |
| Clustering-related | | | |
| Based on internal structure | | | |
| Attributes/variables | | | |
| | Within-cluster homogeneity regarding defining characteristics? | ☐ | |
| | Between-cluster differences account for correlations between variables? | ☐ | |
| (Dis)similarities | | | |
| | Large within-cluster similarities between cluster members? | ☐ | ………………………… |
| | Large similarities between cluster members and cluster-representative object? | ☐ | ………………………… |
| | Small between-cluster similarities? | ☐ | ………………………… |
| Distribution in data space | | | |
| | Cluster connectedness? | ☐ | |
| | Cluster shape? | ☐ | ………………………… |
| | Each cluster concentrated on low dimensional manifold? | ☐ | |
| | Type of between-cluster separability? | ☐ | ………………………… |
| | Cluster centroids in low dimensional subspace? | ☐ | |
| | Cluster area has high density? | ☐ | |
| | Cluster area is surrounded by near-zero density? | ☐ | |
| Based on relation with external variables | | | |
| | Within-cluster homogeneity? | ☐ | ………………………… |
| | Between-cluster differences? | ☐ | ………………………… |

**Part 2: Data-analytic aims (ctd)**

| *Aspect/Sub-aspect* | *Question* | *Check* | *Specification* |
|---|---|---|---|
| Related to quality of data analysis | | | |
| Input preprocessing | | | |
| | Operations needed to address user concerns included? | ☐ | ............................ |
| | Domain knowledge adequately factored in? | ☐ | ............................ |
| Core data analysis | | | |
| | Operations needed to address user concerns included? | ☐ | ............................ |
| | Domain knowledge adequately factored in? | ☐ | ............................ |
| | Quality of algorithmic performance? | ☐ | ............................ |
| Output postprocessing | | | |
| | Operations needed to address user concerns included? | ☐ | ............................ |
| | Domain knowledge adequately factored in? | ☐ | ............................ |
| | Parsimony (Ockham's razor)? | ☐ | ............................ |
| Relation between input and output | | | |
| | Goodness of fit? | ☐ | ............................ |
| | Appropriate influence of different parts of the data on the output? | ☐ | ............................ |
| | Stability or replicability? | ☐ | ............................ |

**Part 3: Data**

| *Aspect/Sub-aspect* | *Question* | *Check* | *Specification* |
|---|---|---|---|
| Structure of data | | | |
| | What are the different sets of elements that are involved in a single data entry? | | ............................ |
| | What kind of combination of the sets is involved in a single data entry? | | ............................ |
| | Are data entries in principle available for all possible combinations of elements? | □ | ............................ |
| | Are some data modes structured in terms of prespecified characteristics? | □ | ............................ |
| | Do the data comprise information of different kinds on the same set of entities? | □ | ............................ |
| Values of data entries | | | |
| | Are the data entries single values? | □ | ............................ |
| | In case of single numeric values: do all numeric relations/operations make sense? | □ | ............................ |
| | Which data entries can be meaningfully compared across the data array? | | ............................ |
| Data availability | | | |
| | Are all data available at once? | □ | ............................ |
| Prior membership info | | | |
| | Is prior information on class membership available? | □ | ............................ |
| | If prior information is available, is it fully available? | □ | ............................ |
| Data-related challenges | | | |
| | Are the data large / ill-conditioned? | □ | ............................ |
| | Are the data sparse? | □ | ............................ |
| | Do the data contain outliers? | □ | ............................ |
| | Are there (many) missing values? | □ | ............................ |
| | Are there any problems with data quality? | □ | ............................ |

**Part 4: Input preprocessing**

| *Aspect/Sub-aspect* | *Question* | *Check* | *Specification* |
|---|---|---|---|
| Basis | | | |
| | Internal data? | ☐ | ............................ |
| | External data? | ☐ | ............................ |
| Level | | | |
| Involvement of data parts | | | |
| | Retaining vs. dropping? | ☐ | ............................ |
| | Weighting? | ☐ | ............................ |
| Changes within initial data structure | | | |
| | Explicitly or implicitly replacing values in (parts of) the data? | ☐ | ............................ |
| | Addition of parts to the data? | ☐ | ............................ |
| Modification of initial data structure | | | |
| | Is a modification of the initial data structure at stake? | ☐ | ............................ |

**Part 5: Core data analysis**

| *Aspect/Sub-aspect* | *Question* | *Check* | *Specification* |
|---|---|---|---|
| Clusters' extension (membership) | | | |
| | What is the nature of the elements of the clusters? | ◻ | ............................. |
| | Is cluster membership crisp or fuzzy? | ◻ | ............................. |
| | Does the method aim at one/a few clusters or at a clustering to be assessed as a whole? | ◻ | ............................. |
| | Are all elements to be clustered? | ◻ | ............................. |
| | Are overlapping clusters possible? | ◻ | ............................. |
| | Does the method aim at a single clustering or at multiple clusterings? | ◻ | ............................. |
| | Do (intrinsic or extrinsic) extensional constraints apply? | ◻ | ............................. |
| | Are they constraints on what may be admissible clusters? | ◻ | ............................. |
| | Are they constraints on what may be admissible clusterings? | ◻ | ............................. |
| Clusters' intension (meaning) | | | |
| | Does each cluster go with pattern of attribute-values/centroid/prototype/representative object(s)? | ◻ | ............................. |
| | Does each cluster go with a cluster membership rule in terms of characteristics or attributes? | ◻ | ............................. |
| | Does each cluster go with a cluster-specific weight that represents feature importance or salience? | ◻ | ............................. |
| | Does each cluster go with a cluster-specific model? | ◻ | ............................. |
| | Do intensional constraints apply? | ◻ | ............................. |

**Part 6: Output postprocessing**

| *Aspect/Sub-aspect* | *Question* | *Check* | *Specification* |
|---|---|---|---|
| Target | | | |
| | Is model selection at stake? | ☐ | ............................. |
| | Is model evaluation at stake? | ☐ | ............................. |
| | Is output interpretation at stake? | ☐ | ............................. |
| Basis | | | |
| | Is target to be addressed on the basis of internal or external information? | ☐ | ............................. |
| | Is target to be addressed on the basis of transformed data-analytic output? | ☐ | ............................. |
| Tools | | | |
| | Is target to be addressed by means of visualization/graphical tools? | ☐ | ............................. |
| | Is target to be addressed by means of numerical tools? | ☐ | ............................. |